\documentclass[12pt, a4paper]{article}

\usepackage[
  margin=0.75in,
  headsep=10pt, 
]{geometry}

\usepackage{graphicx}
\usepackage{soul}
\usepackage{graphicx}
\usepackage{epstopdf, epsfig}
\usepackage{amsmath}
\usepackage[table]{xcolor}
\usepackage{subcaption}
\usepackage{soul}
\usepackage{makecell}
\usepackage{multirow}
\usepackage{breqn}
\usepackage{amssymb} 
\usepackage{booktabs}
\usepackage{dcolumn}
\usepackage{bm}
\usepackage[utf8]{inputenc}
\usepackage[T1]{fontenc}
\usepackage{mathptmx}
\usepackage{etoolbox}
\usepackage{hyperref}
\usepackage{subcaption}
\usepackage{ragged2e}
\usepackage[sort&compress]{natbib} 
\usepackage[utf8]{inputenc}
\usepackage[T1]{fontenc}
\usepackage{graphicx}
\usepackage{epstopdf, epsfig}
\usepackage{xcolor}
\usepackage[table]{xcolor}
\usepackage{soul}
\usepackage{makecell}
\usepackage{multirow}
\usepackage{booktabs}
\usepackage{dcolumn}

\newcommand{\edt}[1]{{\color{black}#1}} 
\soulregister{\edt}{1}
\newcommand{\rev}[1]{{\color{black}#1}} 
\soulregister{\rev}{1}
\newcommand{\af}[1]{{\color{black}#1}} 
\newcommand{\rmv}[1]{{\color{black}#1}} 
\soulregister{\rmv}{1}

\usepackage{amsmath}   
\usepackage{amssymb}   
\usepackage[normalem]{ulem}

\newcommand{\soptitle}{A Fully Parallel Dual-Grid Immersed-Boundary Framework for Flow-Induced Sound from Complex Moving and Deforming Bodies}
\usepackage{xcolor}

\begin{document}

\begin{center}
\Large \bf{\soptitle}
\vspace{0.1in}
\end{center}

\begin{center}
{Amirhossein Fardi and Muhammad Saif Ullah Khalid$^{\star}$}
\vspace{0.1in}
\end{center}
\begin{center}
Nature-Inspired Engineering Research Lab (NIERL)\\ 
Department of Mechanical and Mechatronics Engineering\\
Lakehead University, Thunder Bay, ON P7B 5E1, Canada\\
\vspace{0.05in}
$^\star$\small{Corresponding Author, Email: mkhalid7@lakeheadu.ca}
\end{center} 

\begin{abstract}
Predicting flow-induced sound from moving and deforming bodies is computationally demanding because the near-field hydrodynamics and the far-field acoustics require substantially different spatial resolutions and domain extents. A fully parallel hybrid framework is developed to address this disparity by coupling an incompressible Navier--Stokes solver to an acoustic perturbation equation (APE) solver on independently generated, non-conforming Cartesian grids. A sharp-interface ghost-cell immersed boundary method, with radial-basis-function reconstruction, imposes the boundary conditions for complex moving geometries on both grids. The converged flow field supplies the acoustic source through a one-way, precomputed parallel interpolation operator. This arrangement confines the flow grid to the body and wake while allowing the acoustic grid to extend independently into the far field. The framework is validated for Gaussian-pulse propagation, pulse scattering by a rigid cylinder, tonal sound from flow past a cylinder, and radiation from a traveling wavy foil. The predicted waveforms, wavelengths, pressure amplitudes, and radiation patterns agree closely with analytical solutions and published reference data. Applications to eel and Jack fish locomotion, a four-eel school, a manta ray, and a harbor seal further demonstrate the treatment of realistic three-dimensional morphologies, large boundary deformation, and multiple interacting swimmers. The results resolve morphology-dependent acoustic signatures and interference-driven changes in far-field directivity without requiring the flow grid to span the acoustic far field.
\end{abstract}

\section{Introduction}
\label{sec:Intro}

\edt{Pressure fluctuations are produced by} unsteady \edt{flows} around a wide range of engineering systems, \edt{ranging} from \edt{aircrafts} and underwater vehicles to turbomachinery\edt{. Examining and} controlling the noise radiated \edt{through} these fluctuations \edt{became} an increasingly critical design constraint \cite{colonius2004computational}. \edt{Contrarily}, aquatic and aerial animals achieve a remarkable combination of high propulsive \edt{performance}, low acoustic signature, and long endurance \cite{fish2020bio,fish2020advantages}, which \edt{motivates} research into the mechanisms \edt{responsible for} bio-inspired locomotion \edt{generating} and \edt{propagating} sound. Understanding these mechanisms is essential for developing quiet or acoustically stealthy vehicles, including micro air vehicles and unmanned underwater \edt{robots} \cite{marcin2020fish}. Beyond engineering applications, the sound generated by flapping or undulatory motion is also believed to function as a signal for communication, mating, and predator-prey interactions, thereby influencing the behavior of fish and other animals \cite{seo2021mosquitoes,short2020influence}.

\edt{Experimentally investigating and characterizing underwater noise around stationary and moving bodies is very challenging due to potential contamination from other sources of noise and reflection of sound waves from different objects and boundaries \cite{zhou2024effect}. This situation makes physics-based computational simulations a viable solution for examining flow-induced acoustic signatures around different objects.} Computational approaches for predicting flow-generated sound can be broadly divided into direct and indirect or hybrid methods \cite{wang2006computational}. Within \edt{these} general \edt{frameworks}, three commonly used strategies are direct noise computation, acoustic analogy , and hydrodynamic/acoustic splitting. In \edt{the} direct noise computation, the compressible Navier-Stokes equations are solved to obtain the unsteady flow and radiated acoustic fields simultaneously. This approach can, in principle, describe both sound generation and propagation without introducing a separate acoustic model. The need to maintain the requirements for this approach makes direct computation prohibitively expensive for many practical engineering applications. Indirect methods reduce computational burden by separating the calculation of the unsteady flow from that of the radiated sound. In acoustic analogy\edt{-based} approaches, the flow field is first obtained using conventional computational fluid dynamics\edt{-based} methods, after which the relevant flow quantities are introduced as source terms in a separate acoustic formulation. Representative formulations include Lighthill’s acoustic analogy \cite{lighthill1952sound}, Curle’s extension for flows interacting with stationary solid boundaries \cite{curle1955influence}, the Ffowcs Williams-Hawkings (FWH) equation \af{\cite{williams1969sound}}, and Kirchhoff-type formulations extended to moving surfaces \cite{farassat1988extension}. These methods \edt{were} applied successfully to a wide range of aeroacoustic problems \af{\cite{farassat1998acoustic, lyrintzis2003surface, colonius2004computational, wang2006computational}}. Nevertheless, commonly used free-field implementations often rely on assumptions\edt{,} such as a stationary or uniform propagation medium. Under such assumptions, the effects of a spatially varying background flow on acoustic convection and refraction may not be fully represented, particularly when the source region is non-compact \cite{wang2006computational,schoder2019hybrid}. These restrictions arise from the assumptions adopted in particular analogy\edt{-based} formulations or implementations rather than \edt{originating} from the exact Lighthill framework itself, whose source terms can, in principle, contain effects associated with generation, convection, refraction, and dissipation \edt{of sound}.

The hydrodynamic/acoustic splitting approach provides another form of hybrid computation in which the flow and acoustic disturbances are represented and solved separately. An early formulation was introduced by Hardin and Pope \cite{hardin1994acoustic} and subsequently \edt{further} developed by Shen et al. \cite{shen1999aeroacoustic,shen2004collocated}. \af{In this framework, each primitive flow variable, such as density, velocity, and pressure, is expressed as the sum of a base-flow component and a perturbation component.} The underlying hydrodynamic field is first determined, and the acoustic perturbations are then obtained from a separate system of equations driven by source terms, \edt{which are} derived from the \edt{solutions for the flow dynamics}. Because the acoustic equations may retain the influence of the spatially varying base flow, this approach can describe the convection and refraction of acoustic disturbances in non-uniform flows. It is\edt{,} therefore\edt{,} particularly suitable for low-Mach-number configurations\edt{,} involving spatially distributed or non-compact sources.

\af{Following the flow-acoustic splitting approach outlined in the preceding paragraph,} Ewert and Schröder \cite{ewert2003acoustic} developed the acoustic perturbation equations (APE), in which non-acoustic perturbation components, including vortical and entropy-related disturbances, are filtered from the propagated acoustic field. This treatment suppresses the amplification of hydrodynamic instabilities and improves numerical robustness when acoustically unstable or strongly sheared base flows are considered. \edt{On the contrary}, the linearized Euler equations retain acoustic, vortical, and entropy modes within the same perturbation system \af{, thereby allowing unwanted hydrodynamic instability waves to enter the computed acoustic field.} Seo and Moon \cite{seo2006linearized} proposed the linearized perturbed compressible equations (LPCE) using a similar hydrodynamic/acoustic decomposition, with the objective of isolating the acoustic response while minimizing contamination from perturbed vorticity and hydrodynamic instabilities.

\edt{It becomes substantially challenging with all the} three families \edt{of the afore-mentioned computational techniques} when the \edt{source of} sound is a body undergoing large displacement or deformation, because a body-fitted mesh of adequate quality is then expensive to build and must be regenerated as the body moves. The immersed boundary method (IBM), originally introduced by Peskin \cite{peskin1972flow} to simulate blood \edt{flows} around valves \edt{in a heart}, provides an \edt{alternate} framework\edt{,} in which the governing equations are discretized on a stationary Cartesian grid\edt{, whereas} the solid boundary is represented independently of the underlying mesh. Consequently, mesh regeneration is \edt{not required at all} when the immersed body moves or deforms. This \edt{feature makes} the IBM particularly attractive for flows \edt{interacting with} complex and moving \edt{volumetric and membranous structures}, {the examples of which also include a wide variety of bio-inspired and biological systems} \cite{mittal2005immersed}. More recently, IBM\edt{-based} formulations \edt{were also} extended to acoustic-wave propagation and \edt{prediction of} flow-induced sounds.

\edt{With this context,} Komatsu et al. \cite{komatsu2016direct} and Hattori and Komatsu \cite{hattori2017mechanism} coupled a volume-penalization IBM with \edt{compressible flow-based} direct numerical simulation to investigate \edt{aeroacoustics} generated around stationary and oscillating cylinders. Wang et al. \cite{wang2020immersed} developed a diffusion-interface penalty method for fluid-structure-acoustic interactions involving large structural deformations. Their framework combined a finite difference method\edt{-based flow}  solver with a finite element \edt{technique-based} structural solver\edt{. They} employed a hybrid parallelization strategy based on OpenMP for shared-memory processing and Message Passing Interface (MPI) for communication between computational nodes. Cheng et al. \cite{cheng2021semi} subsequently implemented a semi-implicit, body force-based IBM for \edt{problems related to} viscous flow-induced sound \edt{around} moving bodies. In these diffuse-interface or forcing-based formulations, the boundary condition \edt{for a solid body} is imposed over a finite region surrounding the interface rather than directly at a sharply defined surface.

An \edt{alternate} category of immersed\edt{-}boundary formulations uses ghost cells placed within the solid region to impose the boundary conditions \edt{for solid structures}. Unlike diffuse interface\edt{-based} approaches, ghost-cell methods preserve a sharp representation of the fluid-solid interface. This feature is especially important in computational aeroacoustics, where excessive smoothing near the boundary can introduce numerical dissipation, dispersion, and errors in the amplitude or phase of the propagated acoustic waves. Seo and Mittal \cite{seo2011high} developed a high-order ghost-cell method for the linearized perturbed compressible equations around complex moving bodies for low Mach\edt{-}numbers \edt{flows}. \edt{Furthermore,} Xie et al. \cite{xie2020sharp} employed \edt{the} constrained moving-least-square interpolation to reconstruct ghost-cell values within a linearized Euler equation\edt{-based} framework and demonstrated second-order spatial accuracy for acoustic scattering by geometrically complex rigid bodies. \edt{Additionally,} He et al. \cite{he2022improved} proposed a wavelet-based IBM for fluid-structure-acoustic interaction with elastic boundaries and investigated how boundary stiffness \edt{modified} the resulting acoustic field. However, their \edt{work} was limited to elastic structures anchored at fixed locations and did not consider \edt{their} oscillatory motion, large displacement, or substantial \edt{structural deformations}. \edt{Later,} Zhao et al. \cite{zhao2021sharp} coupled a ghost-cell sharp-interface IBM with the acoustic perturbation equations to predict flow-induced sound around complex stationary geometries, including the noise generated by \edt{flows} through \edt{an array for four cylinders}. Building on this framework, Zhao et al. \cite{zhao2024hybrid} developed a two-stage hybrid method in which an incompressible flow solver was coupled with an acoustic perturbation equation solver, while a ghost-cell sharp-interface IBM was \edt{used} consistently in both computational stages. The method was demonstrated for \edt{aeroacoustic applications involved with} several moving \edt{objects}, including rotating and oscillating cylinders, tandem arrangements of stationary and oscillating cylinders, flapping \edt{wings of insects}, and \edt{undulating} foils \edt{in tandem arrangements}.

\edt{Nevertheless, employing} the same computational grid for both the flow and acoustic solvers presents an important challenge. Although the numerical formulations do not inherently require the flow and acoustic fields to share a single grid \cite{zhao2024hybrid}, doing so forces the extent and resolution of the flow domain to satisfy more demanding \edt{and strict} requirements \edt{related to propagation of} acoustic \edt{waves}. \edt{Particularly}, maintaining low numerical dispersion and dissipation requires a sufficient number of grid points per acoustic wavelength over a large propagation distance. \edt{For simulations for flows with low Mach numbers, this requirement forces to extend the} computational domain to approximately $200-240$ \edt{times the body-based length scales} from \edt{an} immersed body, \edt{which is} far beyond the region required to resolve the \edt{flow features near a body} and wake \edt{dynamics}.


Outside this ghost-cell IBM\edt{-based} family \edt{of computational solvers}, several hybrid \edt{computational aero-acoustic (CAA)} solvers already avoid a single shared grid. Smith and Ventikos \cite{smith2022hybrid} \edt{computed} the \edt{solutions for flows and acoustics} together\edt{,} rather than \edt{doing it sequentially} on two grids that \edt{overlapped partially}. \edt{Their approach involved carrying} the flow-\edt{related} source terms across to the \edt{grid for the acoustic solver} through a radial-basis-function interpolation\edt{. Here,} advancing both fields \edt{simultaneously} means the flow \edt{parameters} do not have to be \edt{recorded} and reloaded at every step, though the two meshes still cover much of the same region. \edt{Moreover,} Purohit et al. \cite{purohit2014numerical} \edt{kept} the flow- and the acoustic-solution steps sequential and place the acoustic sources on an enclosing surface rather than throughout a volume, with the acoustic mesh spaced and stepped differently from the flow mesh and an interpolation applied between the two \edt{fields}. \edt{For large-eddy simulations (LES),} Labbé et al. \cite{labbe2013cfd} \edt{divided} a jet-noise problem into a small inner region \edt{discretized} with a structured mesh and a larger outer region \edt{handled} with an unstructured discontinuous-Galerkin mesh\edt{. The two grids were} joined at a buffer zone\edt{,} where \edt{the flow-related} data \edt{was} passed outward at every step through an interpolation onto the quadrature points of the acoustic mesh\edt{. T}he two \edt{grids shared a nonconformal interface}, and \edt{the solution on the outer grid was computationally more expenseive in terms of the required time while} locally \edt{adjusting} time steps. \edt{Besides,} Gröschel et al. \cite{groschel2008noise} \edt{obtained} a jet flow field by LES and\edt{,} then\edt{, evaluated} the far-field sound on a second, coarser grid\edt{, which extended} past the flow domain. They also showed that the APE\edt{-based} solution \edt{was} less sensitive to some source-region truncations \edt{compared with that from FWH technique}. \edt{Moreover,} Moon et al. \cite{moon2010hybrid} noted that \edt{separation of grids could} \edt{greatly help with dealing with strict constraints on the timestep size through the Courant-Friedrichs-Lewy (CFL) criterion as well as reducing} the flow domain.

\edt{With this background, our present work addresses the following important limitations and gaps in the existing computational techniques for dealing with flow-induced noise: (i) for low-Mach number flows, such as the ones in biological swimming, characteristic swimming velocities are considerably smaller than the speed of sound in water, and hydrodynamic pressure fluctuations associated with the body's motion and wake can be substantially stronger than the relatively weak acoustic perturbations, making accurate prediction of sound generation and propagation particularly challenging on a traditionally used single computational grid, (ii) existing solvers are often limited in their ability to accommodate flexible bodies and to represent both volumetric and membranous structures, undergoing large-amplitude oscillations, within a unified computational formulation, (iii) most of the currently available computational approaches rely on low-order acoustic models and acoustic analogies, that primarily estimate the radiated acoustic field without directly resolving the coupled flow-acoustic dynamics, thereby limiting the accurate identification and physical interpretation of sound-generation mechanisms that arise from vortex-vortex interactions, vortex-body interactions, and unsteady loading on stationary, moving, or deforming surfaces. It is important to mention that the flow-induced noise from these underwater sources are important sensory cues potentially to be used for underwater perception and navigation.} \edt{Therefore, these specific research questions frame the \edt{objectives and} novelty of our current research. This work advances the computational technique based on the ghost-cell sharp-interface IBM, recently presented by Farooq et al. \cite{farooq4874977accurate} related to incompressible fluid-structure interactive systems and introduces a two-independent grids strategy for solving the APE-based formulation within the fully parallelized framework. There exists one-way coupling between the flow and acoustic grids,} where the flow is solved to convergence first, and the flow solution and the acoustic source terms are then interpolated from the flow grid onto the acoustic grid at every flow time step. Because the two grids are decoupled, the flow grid can be confined to the region needed to resolve the wake of the moving \edt{bodies, whereas} the acoustic grid is \edt{constructed} separately \edt{to satisfy the conditions for the required number of grid points per wavelength} of the high-order scheme over the larger domain needed for far-field directivity\edt{. It is important to note that these measures are taken here} without enlarging the flow domain to match it. Both the flow and the acoustic \edt{solvers} are also fully parallelized with the open message passing interface (MPI) through domain decomposition\edt{.} \edt{It} allows larger and more highly resolved grids to be used within a \edt{more practically meaningful} computing time. Together, the dual-grid coupling and the parallel implementation are intended to reduce the cost of the flow \edt{computations} and to \edt{have more freedom to extend and/or refine} the acoustic grid, whenever it is needed, while extending the two-step ghost-cell IBM approach to larger and more demanding moving-boundary problems\edt{, such as fish schooling or bird flocking configurations}.





\section{Computational Methodology}
\label{sec:comp_meth}


The flow field that \edt{serves as} the acoustic sources is obtained from fully three-dimensional ($\mbox{3D}$) numerical simulations. \edt{Here, motion} of the fluid is described by the nondimensional \edt{incompressible} forms of the continuity and Navier--Stokes equations, which are:

\begin{equation}
\frac{\partial u_j}{\partial x_j} = 0
\label{eqn:continuity}
\end{equation}

\begin{equation}
\frac{\partial u_i}{\partial t}
+ u_j \frac{\partial u_i}{\partial x_j}
=
-\frac{1}{\rho}\frac{\partial p}{\partial x_i}
+ \frac{1}{\mathrm{Re}}
\frac{\partial^2 u_i}{\partial x_j \partial x_j}
\label{eqn:momentum}
\end{equation}

\noindent where the indices satisfy $\{i,j\}=\{1,2,3\}$, $x_i$ denotes the Cartesian coordinate directions, $u_i$ are the Cartesian components of \edt{velocity of the} the fluid, $p$ is \edt{the flow} pressure, and $\mbox{Re}={U_\infty}{L}/\nu$ is the Reynolds number\edt{, where $U_\infty$ and $\nu$ present the flow velocity and kinematic viscosity of the fluid. Here, $L$ denotes the length scale of the body placed in the fluid flow, and $\rho$ is density of the fluid.}

These governing equations are \edt{through} a sharp-interface IBM formulated on a non-uniform Cartesian grid, in which radial-basis functions are \edt{employed} as the interpolation scheme for the boundary, so that the immersed bodies are represented accurately \cite{farooq4874977accurate}. The IBM provides an effective computational framework for simulating the complex fluid-structure interactions that arise in demanding configurations, such as those associated with the biological swimmers considered in this \edt{work}. Being a practical and physically consistent strategy for investigating flows that interact with biological systems\edt{,} such as swimming fish, the IBM is particularly well suited \edt{for} handling large deformations \edt{of volumetric and/or membranous structures}, and intricate geometrical configurations that are difficult to treat with conventional body-fitted mesh-based methods. By embedding complex geometries within a Cartesian grid, the IBM separates the mesh generation from the geometry of the body, which greatly simplifies the mesh-generation process and removes the need for computationally expensive remeshing or grid deformation at every time step \cite{mittal2008versatile, farooq4874977accurate}. This separation makes it straightforward to \edt{computationally deal with} complex geometries and substantially lowers both the cost of preprocessing immersed rigid or flexible structures and the associated computational overhead. In addition, the ability of the method to represent infinitesimally thin structures, such as fins, broadens its applicability to the wide range of morphologies exhibited by aquatic swimmers \cite{farooq4874977accurate}.

\edt{In Eq.~\ref{eqn:momentum},} spatial discretization of the diffusion term is carried out with a central difference scheme, whereas the convection term is discretized using the Quadratic Upstream Interpolation for Convective Kinematics (QUICK) scheme. \edt{Integration in time} relies on a fractional-step method and achieves second-order accuracy in both time and space. \edt{The presence or kinematics of a structural object} is imposed as a boundary condition \edt{its body} and is enforced through a ghost-cell technique that is applicable to both rigid and flexible structures \cite{farooq4874977accurate}. Neumann boundary conditions are prescribed at the far-field boundaries, with the exception of the inlet boundary, where Dirichlet conditions are imposed to specify the inflow. Further details of this fully parallelized solver \edt{along} with its application to a variety of problems governed by complex fluid-structure interactions can be found in Ref. \cite{farooq4874977accurate}.


\edt{The linearized Euler Equations ($\mbox{LLE}$) provide} a widely \edt{adopted} framework for modeling how sound travels through a spatially varying background flow \cite{zhao2021sharp}. \edt{It is because} they inherently capture convection and refraction that the mean flow imposes on an acoustic disturbance. A drawback of this framework is that its solution space is not restricted to the acoustic mode alone\edt{.} \edt{These formulations} simultaneously carry vortical and entropic disturbances \cite{zhao2021sharp}. Whenever \edt{some} forcing excites these additional modes, \edt{the solution for} LEE may grow \edt{unboundedly} and swamp the physical sound field that \edt{is to be computed}. \edt{Such} a situation is especially likely \edt{to happen} when the background field is prone to hydrodynamic \edt{instabilities}. A convenient \edt{method to eradicate} this difficulty was introduced by Ewert and Schr\"oder \cite{ewert2003acoustic}, who recast the propagation model under the assumptions that the acoustic disturbance \edt{was} both curl-free and isentropic. \edt{With these considerations,} the vortical and entropic \af{terms} are filtered out of the \edt{governing} system altogether, and a reduced APE system \edt{remains} that transports only the acoustic mode. Forcing of these equations is \edt{modeled through the} source terms that are reconstructed from the background flow. \edt{Using compact} vector notation, the APE system used here takes the \edt{following} form\edt{:}

\begin{equation}
\frac{\partial p'}{\partial t}
+ \bar{c}^{\,2}\,\nabla\cdot\!\left(\bar{\rho}\,\mathbf{u}^{a}
+ \bar{\mathbf{u}}\,\frac{p'}{\bar{c}^{\,2}}\right)
= \bar{c}^{\,2}\,S_{\mathrm{cont}},
\label{eqn:APEcont}
\end{equation}

\begin{equation}
\frac{\partial \mathbf{u}^{a}}{\partial t}
+ \nabla\!\left(\bar{\mathbf{u}}\cdot\mathbf{u}^{a}\right)
+ \nabla\!\left(\frac{p'}{\bar{\rho}}\right)
= \mathbf{S}_{\mathrm{mom}},
\label{eqn:APEmom}
\end{equation}

\noindent \edt{where} $p'$ stands for the fluctuating acoustic pressure and $\mathbf{u}^{a}$ for the curl-free acoustic velocity. \edt{Also,} $\bar{p}$, $\bar{\rho}$, $\bar{\mathbf{u}}$\edt{,} are the mean pressure, density, \edt{and} velocity \edt{of the flow, respectively with $\bar{c}$ the speed of sound} obtained by time-averaging. \edt{In this manuscript}, a prime marks a fluctuating quantity\edt{,} and an overbar \edt{presents} a time-averaged one. The first relation, Eq.~\eqref{eqn:APEcont}, plays the role of the continuity equation for the acoustic field\edt{,} and the second one, Eq.~\eqref{eqn:APEmom}, \edt{is analogous to} the \edt{conservation of momentum in flows}. Everything on the left of these two equations governs how the sound is carried and bent by the inhomogeneous mean flow, while the quantities $S_{\mathrm{cont}}$ and $\mathbf{S}_{\mathrm{mom}}$ on the right supply the acoustic forcing\edt{,} entering the continuity and momentum balances.

For the external, low-Mach-number flow problems considered here, the source terms \edt{gets simplified}. The momentum source is evaluated directly from the incompressible flow field as\edt{:}

\begin{equation}
\mathbf{S}_{\mathrm{mom}} \simeq \frac{\nabla P'}{\rho_{0}}
= \frac{\nabla\!\left(P-\bar{P}\right)}{\rho_{0}} \edt{;}
\\\\
{S_{\mathrm{cont}} = 0}
\label{eqn:APEsource}
\end{equation}

This expression \edt{precisely} links the two solvers\edt{, where} pressure fluctuations resolved on the flow grid are \edt{transmitted} as the forcing \edt{for} APEs on the acoustic grid. In \edt{the} acoustic simulations, accurate prediction of wave propagation requires numerical methods capable of maintaining characteristics \edt{of the waves} with minimal dispersion and dissipation over long distances. This requirement is typically \edt{fulfilled} through the use of high-order finite-difference schemes. For this purpose, Tam and Webb \cite{tam1993dispersion} developed a seven-point, fourth-order dispersion-relation-preserving (DRP) scheme that significantly \edt{reduced} both dispersion and dissipation errors. For temporal discretization, high-order Runge-Kutta (RK) methods are commonly employed. Building upon this framework, Hu et al. \cite{hu1996low} optimized the RK coefficients and proposed a low-dissipation and low-dispersion Runge-Kutta (LDDRK) scheme, \edt{which further improved} the accuracy of time integration by minimizing numerical errors.

For a uniformly spaced mesh, the DRP approximation of the first derivative on its symmetric stencil is expressed as

\begin{equation}
\frac{\partial f}{\partial x}\left(x_{0}\right)
\simeq \frac{1}{\Delta x}\sum_{j=-3}^{3} a_{j}\,f\!\left(x_{0}+j\,\Delta x\right),
\label{eqn:DRP}
\end{equation}

\noindent with the weights $a_{j}$ given in Table~\ref{tab:DRP} \cite{tam1993dispersion,tam1995computational}. Close to the solid wall, where the full seven-point stencil no longer fits, a conventional central-difference approximation takes over.

\begin{table*}[h]
\centering
\caption{Coefficients of the dispersion-relation-preserving scheme $\left(a_{-j}=-a_{j}\right)$.}
\label{tab:DRP}
\begin{tabular}{cccc}
\hline
$a_{0}$ & $a_{1}$ & $a_{2}$ & $a_{3}$ \\
\hline
$0$ & $0.77088238051822552$ & $-0.166705904414580469$ & $0.02084314277031176$ \\
\hline
\end{tabular}
\end{table*}

Because the present solver operates on a stretched, non-uniform Cartesian mesh, the derivative cannot be taken directly with the uniform-grid stencil without losing accuracy. The remedy is to map the physical, non-uniform coordinate $x$ onto an arbitrary evenly spaced computational coordinate $\xi$ and to differentiate there,

\begin{equation}
\frac{\partial f}{\partial x}
= \frac{\partial \xi}{\partial x}\frac{\partial f}{\partial \xi}
= \frac{1}{\partial x/\partial \xi}\frac{\partial f}{\partial \xi},
\label{eqn:transform}
\end{equation}

\noindent where $\xi$ is the underlying uniform grid. Evaluating the DRP stencil in this mapped coordinate, the derivative at the $m$th node \edt{is computed as:}

\begin{equation}
\frac{\partial f}{\partial x}\left(x_{m}\right)
\simeq \frac{1}{\widetilde{\Delta x}}\sum_{j=-3}^{3} a_{j}\,f_{m+j},
\qquad \text{with} \qquad
\widetilde{\Delta}x = \sum_{j=-3}^{3} a_{j}\,x_{m+j}
\label{eqn:DRPnonuniform}
\end{equation}

\edt{Here,} the geometric \edt{mapping} metric $\partial \xi/\partial x = 1/\left(\partial x/\partial \xi\right)$ is evaluated with the very same high-order DRP operator to avoids a reduction in the formal order of accuracy on the stretched mesh. The \edt{temporal} evolution of the discretized system can be \edt{presented} in the \edt{following} compact form:

\begin{equation}
\frac{\partial U}{\partial t} = F\left(U\right),
\label{eqn:timeevo}
\end{equation}

\noindent in which $U$ gathers the acoustic unknowns\edt{,} and $F(U)$ represents the assembled spatial operators. Advancement in time is performed with the explicit six-stage LDDRK scheme, marching from level $n$ to level $n+1$ through the \edt{following} sequence\edt{:}

\begin{equation}
\begin{aligned}
U^{0} &= U^{n},\\
U^{l} &= U^{n} + \beta_{l}\,\Delta t\,F\!\left(U^{l-1}\right)
\qquad \text{for } l = 1,\ldots,6\\
U^{n+1} &= U^{6},
\end{aligned}
\label{eqn:LDDRK}
\end{equation}

\noindent with $\Delta t$ \edt{as} the time increment and the stage weights $\beta_{l}$ \edt{provided} in Table~\ref{tab:LDDRK} \cite{hu1996low}.

\begin{table}[]
\centering
\caption{Coefficients of the six-stage low-dissipation and low-dispersion Runge--Kutta scheme.}
\label{tab:LDDRK}
\begin{tabular}{cccccc}
\hline
$\beta_{1}$ & $\beta_{2}$ & $\beta_{3}$ & $\beta_{4}$ & $\beta_{5}$ & $\beta_{6}$ \\
\hline
$0.169193539$ & $0.1874412$ & $1/4$ & $1/3$ & $1/2$ & $1$ \\
\hline
\end{tabular}
\end{table}

Although high-order finite-difference schemes significantly reduce numerical dispersion and dissipation, unresolved grid-to-grid oscillations may still develop during the simulation and contaminate the acoustic solution. To suppress these non-physical high-frequency errors and maintain numerical stability, a tenth-order spatial filter \cite{bogey2004family} is \edt{employed} at each iteration. Near the boundaries, the \edt{order of the} filter is gradually reduced to eighth, sixth, fourth, and second order to accommodate the reduced stencil size. For a symmetric $(2N+1)$-point stencil, the filtering operation is given by:

\begin{equation}
D_{f}\left(x\right) = \sigma_{d}\sum_{j=-N}^{N} d_{j}\,f\!\left(x_{0}+j\,\Delta x\right)
\label{eqn:filter}
\end{equation}

\noindent where the coefficients $d_{j}$ are \edt{provided} in Table~\ref{tab:filter}\edt{,} and the scalar $\sigma_{d}$, chosen between $0$ and $1$, controls how aggressively the filter acts.

\begin{table*}[h]
\centering
\caption{Coefficients of the filters $\left(d_{-j}=d_{j}\right)$.}
\label{tab:filter}
\begin{tabular}{lcccccc}
\hline
 & $d_{0}$ & $d_{1}$ & $d_{2}$ & $d_{3}$ & $d_{4}$ & $d_{5}$ \\
\hline
10th order & $63/256$ & $-105/512$ & $15/128$ & $-45/1024$ & $5/512$ & $-1/1024$ \\
8th order  & $35/128$ & $-7/32$   & $7/64$   & $-1/32$   & $1/256$ & \\
6th order  & $5/16$   & $-15/64$  & $3/32$   & $-1/64$   &         & \\
4th order  & $3/8$    & $-1/4$    & $1/16$   &          &         & \\
2nd order  & $1/2$    & $-1/4$    &          &          &         & \\
\hline
\end{tabular}
\end{table*}


\edt{Now, we explain the treatment of the boundary conditions in our computational framework for handling complex fluid-structure-acoustic interactive systems.} Both the flow and the acoustic problems are solved on non-uniform Cartesian grids \edt{here} that do not conform to the geometry's surface\edt{. Therefore,}  the conditions on the body are enforced through the IBM rather than through a body-fitted mesh. Two overlapping descriptions of the geometry coexist: \edt{(i)} the volumetric Cartesian grid that carries the field variables, and \edt{(ii)} an unstructured surface mesh of Lagrangian markers, triangular facets in three dimensions, that discretizes the immersed surface $\Gamma_{b}$ and separates the fluid region $\Omega_{f}$ from the solid region $\Omega_{s}$. On the basis of $\Gamma_{b}$, every Cartesian cell is tagged as a fluid cell, a solid cell, or a ghost cell, the last being a solid cell that has at least one fluid neighbour (north, south, east, west, front, or back). The tagging is obtained from the sign of $\mathbf{p}\cdot\mathbf{n}$, where $\mathbf{p}$ is the position vector of the cell centre relative to the centroid of its nearest surface element and $\mathbf{n}$ is the outward unit normal of that element\edt{. A} positive sign places the cell in the fluid, a negative sign \edt{tags} in the solid. For a stationary body\edt{,} this classification is performed once during pre-processing, whereas it is repeated at every time step for moving \edt{bodies}, as the surface sweeps through the grid. \edt{Then,} the boundary conditions are imposed by prescribing suitable values at the ghost cells through a local reconstruction built on radial-basis-function (RBF) interpolation \cite{farooq4874977accurate}. The treatment differs between the flow field, where viscosity is retained\edt{,} and the no-slip condition \edt{gets applied. Besides,} the medium is taken to be inviscid and a slip condition \edt{is employed in the acoustic field}. \edt{These conditions are formulated and explained now, later followed by elucidating} the \edt{numerical} treatment used at the outer edges of the computational domain.

The incompressible Navier-Stokes equations are subject to the no-slip and no-penetration requirements on the moving surface, so that the fluid\edt{'s} velocity there matches \edt{with} the local velocity of the body\edt{:}

\begin{equation}
\mathbf{u} = \mathbf{u}_{b} \qquad \text{on } \Gamma_{b}
\label{eqn:BCflowvel}
\end{equation}

\noindent where $\mathbf{u}_{b}$ is the prescribed velocity of the \edt{structural} surface, obtained by differentiating the imposed kinematics in time. For pressure, a homogeneous Neumann condition is \edt{employed} in the \edt{direction normal to the wall:}

\begin{equation}
\frac{\partial p}{\partial n} = 0 \qquad \text{on } \Gamma_{b}
\label{eqn:BCflowpres}
\end{equation}

\noindent with $n$ \edt{as} the coordinate along the \edt{locally normal to the wall}. Both conditions are \edt{utilized} through the multi-dimensional ghost-cell variant of the sharp-interface immersed-boundary method \cite{farooq4874977accurate, mittal2008versatile}. For every ghost cell\edt{,} a boundary-intercept (BI) point $\mathbf{x}_{\mathrm{BI}}$ is first located by projecting the ghost-cell center $\mathbf{x}_{\mathrm{GC}}$ normally onto the nearest triangular surface element\edt{. W}hen no normal projection lies inside a surrounding element, the closest element centroid is used instead. An image point (IP) is then defined as the reflection of the ghost cell about the surface, so that the BI point bisects the segment joining the ghost cell to the image point\edt{.}

\begin{equation}
\mathbf{x}_{\mathrm{IP}} = 2\,\mathbf{x}_{\mathrm{BI}} - \mathbf{x}_{\mathrm{GC}}
\label{eqn:BCip}
\end{equation}

\noindent which places the image point in the fluid region close to the surrounding fluid cells. A generic flow quantity $\phi$ (a velocity component or pressure) is reconstructed at the image point by RBF interpolation over a scattered set of $N_{c}$ nearby cell-centered data points $\mathbf{x}_{\mathbf{c}j}$. 

Following the mesh-free spline formulation \edt{explained in} Ref.~\cite{farooq4874977accurate}, the interpolation function is represented as a linear combination of radial basis functions and a low-order polynomial. This formulation yields interpolation weights that depend only on the geometry of the stencil and the image-point location, allowing the image-point value to be evaluated as\edt{:}

\begin{equation}
\phi_{\mathrm{IP}} \approx \sum_{j=1}^{N_{c}} \gamma_{j}\,\phi_{j}
\label{eqn:BCweights}
\end{equation}

\noindent where $\gamma_j$ \edt{represents} the RBF interpolation weights. The value carried by the ghost cell is finally set according to the type of boundary condition. For the Dirichlet velocity condition of Eq.~\eqref{eqn:BCflowvel}, the ghost value is obtained by mirroring the image-point value about the prescribed boundary value $\phi_{\mathrm{BI}}=\mathbf{u}_{b}$\edt{:}

\begin{equation}
\phi_{\mathrm{GC}} = 2\,\phi_{\mathrm{BI}} - \phi_{\mathrm{IP}}
= 2\,\phi_{\mathrm{BI}} - \sum_{j=1}^{N_{c}} \gamma_{j}\,\phi_{j}
\label{eqn:BCdirichlet}
\end{equation}

For the homogeneous Neumann pressure condition of Eq.~\eqref{eqn:BCflowpres}, which enters the pressure Poisson equation, the ghost value is instead extrapolated so that the prescribed normal gradient at the wall is recovered\edt{:}

\begin{equation}
\phi_{\mathrm{GC}} = \phi_{\mathrm{IP}} - \Delta l \left(\frac{\partial \phi}{\partial n}\right)_{\!\mathrm{BI}}
= \sum_{j=1}^{N_{c}} \gamma_{j}\,\phi_{j} - \Delta l \left(\frac{\partial \phi}{\partial n}\right)_{\!\mathrm{BI}}
\label{eqn:BCneumann}
\end{equation}

\noindent where $\Delta l = \|\mathbf{x}_{\mathrm{IP}}-\mathbf{x}_{\mathrm{GC}}\|$ is the length of the ghost-cell-to-image-point segment and $(\partial\phi/\partial n)_{\mathrm{BI}}=0$ for pressure \edt{on the wall}. The weights $\gamma_{j}$ are recomputed once per time step, and Eqs.~\eqref{eqn:BCdirichlet} and \eqref{eqn:BCneumann} are \edt{subsequently used} at every iteration so that the reconstruction stays consistent with the evolving interior field. Because the entire ghost-cell configuration is rebuilt at each step, the method accommodates the large deformations and continuous motion of flexible \edt{complex} geometries without any regeneration of the mesh.


\edt{To explain} the acoustic\edt{s-related computations now,} the medium is regarded as inviscid, so the sound field must satisfy the rigid-wall conditions of vanishing normal acoustic-pressure gradient and vanishing normal acoustic velocity. In terms of the acoustic pressure perturbation $p'$ and the acoustic velocity perturbation $\mathbf{u}^{a}$\edt{,} 

\begin{equation}
\frac{\partial p'}{\partial n} = 0, \qquad \mathbf{u}^{a}\cdot\mathbf{n} = 0
\qquad \text{on } \Gamma_{b},
\label{eqn:BCacoustic}
\end{equation}

\noindent where $\mathbf{n}$ is the outward unit wall-normal vector\edt{. The} first relation \edt{ensures no} normal acoustic-pressure gradient at the wall, while the second \edt{one guarantees zero} flux of the acoustic velocity through it, leaving the tangential motion free. These conditions are enforced at the acoustic-grid ghost cells using the RBF procedure introduced for the flow field. The corresponding BI and IP values are defined as follows\edt{:}

\begin{equation}
p'_{\mathrm{IP}} = \sum_{j=1}^{N_{c}} \gamma_{j}\,p'_{j}, \qquad
\mathbf{u}^{a}_{\mathrm{IP}} = \sum_{j=1}^{N_{c}} \gamma_{j}\,\mathbf{u}^{a}_{j}.
\label{eqn:BCacoustIP}
\end{equation}

The inviscid, slip character of the wall then dictates how the image-point values are transferred to the ghost cell. The homogeneous Neumann condition on the acoustic pressure, $(\partial p'/\partial n)_{\mathrm{BI}}=0$, is the Neumann case of Eq.~\eqref{eqn:BCneumann}, which reduces to a simple mirror of the image-point pressure onto the ghost cell\edt{:}

\begin{equation}
p'_{\mathrm{GC}} = p'_{\mathrm{IP}} = \sum_{j=1}^{N_{c}} \gamma_{j}\,p'_{j}
\label{eqn:BCacoustP}
\end{equation}

\edt{Furthermore,} the no-penetration condition on the acoustic velocity is imposed by reflecting the reconstructed image-point velocity about the wall, that is, by removing twice its wall-normal component\edt{:}

\begin{equation}
\mathbf{u}^{a}_{\mathrm{GC}} = \mathbf{u}^{a}_{\mathrm{IP}}
- 2\left(\mathbf{u}^{a}_{\mathrm{IP}}\cdot\mathbf{n}\right)\mathbf{n}
\label{eqn:BCacoustU}
\end{equation}

\noindent so that the interpolated normal component is inverted across the surface while the tangential component is preserved.

\subsection{Far-field and domain-boundary conditions}
\label{sec:BC_farfield}

At the outer edges of the finite computational domain, distinct treatments are \edt{conducted} for the two solvers. For the incompressible flow, Dirichlet conditions specify the uniform inflow velocity $U_{\infty}$ at the upstream (inlet) boundary, while homogeneous Neumann conditions are imposed on the velocity at the remaining far-field boundaries so that the wake may leave the domain without spurious constraint.

For the acoustic field, the outer boundary must \edt{allow} the radiated sound escape without reflecting it back into the domain, \edt{because} any \edt{reflection from boundaries of the domain} would corrupt the computed wave field. \edt{It} is \edt{attained} with the Energy Transfer and Annihilation (ETA) non-reflecting treatment of Edgar and Visbal \cite{edgar2003general}, \edt{which is} a buffer-zone method \edt{to combine} grid stretching with high-order low-pass filtering. The idea is to first \emph{transfer} the energy of the outgoing waves into progressively higher wavenumber (shorter-wavelength) modes by aggressively coarsening the mesh inside a buffer layer, and then\edt{,} to \emph{annihilate} this high-wavenumber content \edt{that} the interior scheme can no longer propagate \edt{effectively} with a spatial filter before it can reflect. Within the buffer layer\edt{,} the grid is stretched geometrically, so that consecutive spacings grow by a constant ratio\edt{:}

\begin{equation}
\Delta x_{k+1} = C_{\mathrm{ETA}}\,\Delta x_{k}, \qquad k = 1, 2, \dots, N_{\mathrm{ETA}}
\label{eqn:BCstretch}
\end{equation}

\noindent where $C_{\mathrm{ETA}}$ is the stretching (growth) rate\edt{,} and $N_{\mathrm{ETA}}$ \edt{shows} the number of points in the zone. Following the parameter\edt{ic} study of Edgar and Visbal \cite{edgar2003general}, we use an aggressive growth rate of $C_{\mathrm{ETA}}=2$ with $N_{\mathrm{ETA}}=10$ points in the buffer layer along every outer boundary\edt{, presenting} a combination that was shown to reduce the reflected energy to well below one percent of the root-mean-square signal. 

\subsection{Dual-Grid Strategy for Flow-Acoustic Coupling}
\label{sec:decomposition}

The hydrodynamic and acoustic fields are separated by several orders of magnitude in both amplitude and characteristic length scale. \edt{Hence,} resolving the energetic near-field flow while simultaneously propagating the weak acoustic field to the far field on a single mesh is prohibitively expensive\edt{. T}he flow region demands fine resolution near the immersed surface and in the wake, whereas the acoustic region requires only a small number of points \edt{spanning over} an acoustic wavelength. \edt{However, the number of required points for the overall domains becomes huge}. To resolve this conflict\edt{,} we adopt \edt{and report} a dual-grid strategy \edt{in our present work,} in which the incompressible flow and acoustics are advanced on two distinct, independently decomposed Cartesian grids that overlap in \edt{the} physical space and are coupled by a one-way interpolation operator. \edt{Now, we explain} the two principal \edt{novel} contributions of the present work: (i)~the dual-grid decomposition and its coupling, and (ii)~a precomputed, fully parallel trilinear interpolation operator that transfers the acoustic-source data between the two grids at negligible per-step cost.

\begin{figure}
    \centering
    \includegraphics[width=1.0\linewidth]{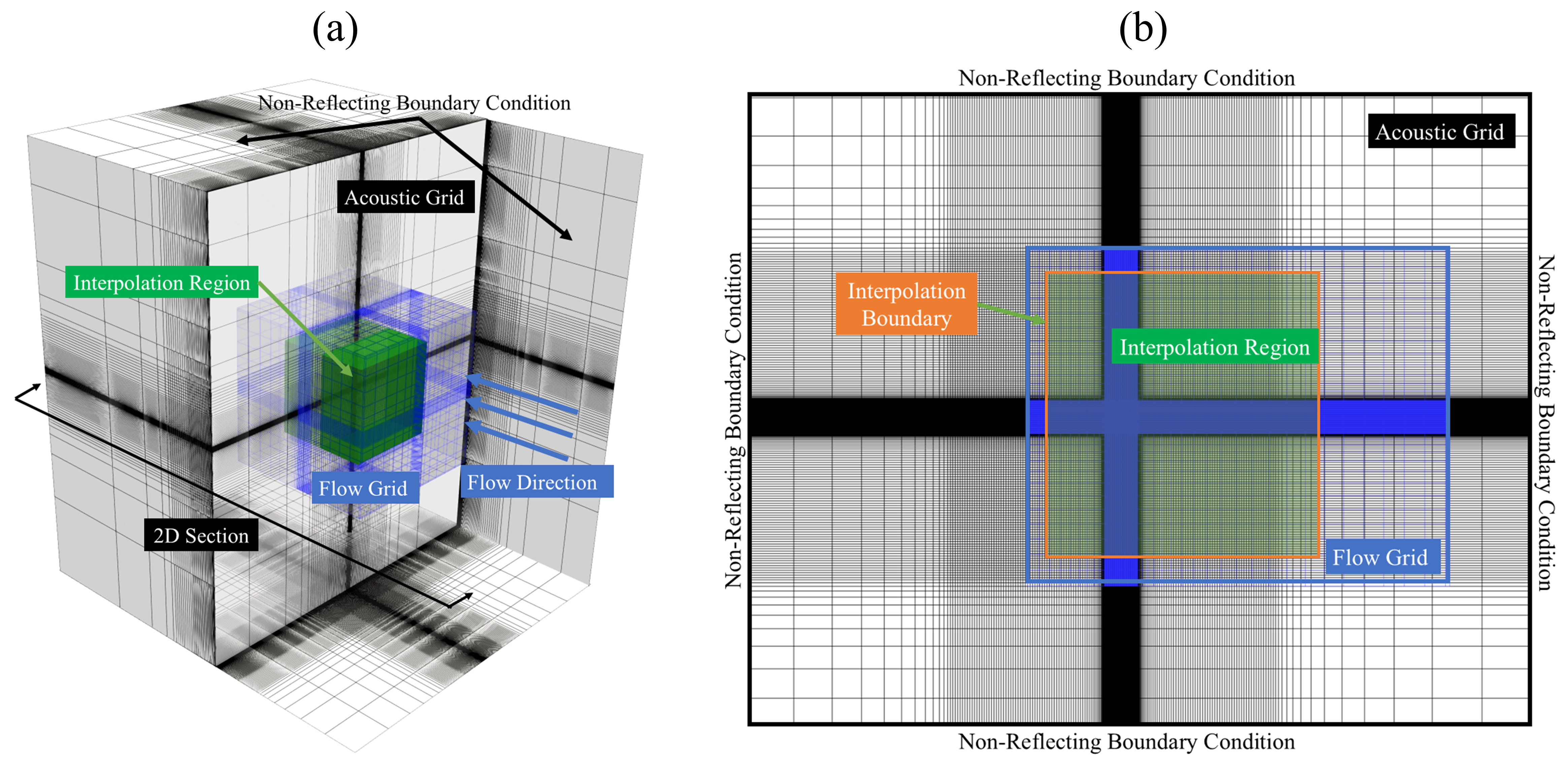}
    \caption{Schematic of the computational domain illustrating the dual-grid configuration, including the flow grid, acoustic grid, and interpolation region. (a) Three-dimensional view. (b) Two-dimensional cross-sectional view including the interpolation boundary.}
    \label{fig:Overlap_Domain}
\end{figure}

\edt{For the first pointer, we} let the flow be solved on a grid $\mathcal{G}_F$ (\emph{flow grid}), and let the acoustic field be solved on a second grid $\mathcal{G}_A$ (\emph{acoustic grid})\edt{, as presented in} Fig.~\ref{fig:Overlap_Domain}. The two grids are generated independently \edt{through} stretched, non-uniform distributions, so that $\mathcal{G}_F$ concentrates points in the near-wall flow region\edt{, whereas} $\mathcal{G}_A$ provides a non-uniform, acoustically adequate resolution over a substantially larger extent. The grids are co-located in \edt{the} physical space,
$\mathcal{G}_F \subseteq \mathcal{G}_A$, but \edt{they} are not required to share nodes, spacing, or block boundaries\edt{. T}he acoustic grid typically extends well beyond the flow grid so that the radiating field can be propagated and damped before it reaches the outer \edt{boundaries. Nevertheless,} the two grids are constructed to share the same minimum spacing \edt{in the immediate vicinity of an} immersed body. Because both solvers must resolve \edt{the geometric structural features} with comparable fidelity, the smallest cell size of the
acoustic grid is matched to that of the flow grid in the near-body region. Where The two node sets are aligned \edt{where the spacings coincide}, and the transfer of data there is not a genuine interpolation but an \emph{exact node-to-node map}
(equivalently, a permutation that may cross process boundaries), so that no
interpolation error is incurred close to the surface where accuracy matters
most.

\edt{For parallelization of this computational framework,} the two grids are decomposed independently across the available processes. Adopting a lexicographic Cartesian partition over a process grid of size $P_x \times P_y \times P_z$, the flow node with global indices $(i,j,k)$ is owned by the rank:

\begin{equation}
  \mathcal{R}_1(i,j,k) \;=\; p_x(i) \;+\; P_x\Big(\,p_y(j) \;+\; P_y\,p_z(k)\Big),
  \label{eq:decomp}
\end{equation}

\noindent where $p_x(i)\in\{0,\dots,P_x-1\}$ is the process column that contains the global index $i$ (\edt{with} $p_y$ and $p_z$ defined analogously), determined from the prefix sums of the per-process block sizes. An identical rule
$\mathcal{R}_2$ governs the acoustic grid with its own process grid. Because
$\mathcal{G}_F$ and $\mathcal{G}_A$ have different global sizes and independent partitions, the same physical point generally belongs to different ranks on the two grids\edt{. A} given rank owns a block of the flow field and a geometrically unrelated block of the acoustic field, and the mapping between them is many-to-many across processes. Handling this mismatch efficiently is the central difficulty
addressed by the interpolation operator in the following section.

\edt{To elaborate the handling of one-way} coupling between the grids, the flow field forces the acoustic field, but the acoustic perturbation does not \edt{provide any} feedback \edt{to} the flow. \edt{When} the flow\edt{-related} statistics are converged, the acoustic source is built from the flow pressure, which is interpolated from $\mathcal{G}_F$ onto $\mathcal{G}_A$ at every time step, exchanged across the acoustic halos, and then advanced in time by the acoustic solver.

\subsection{Precomputed parallel interpolation}
\label{sec:interpolation}

The transfer of the flow data from $\mathcal{G}_F$ to $\mathcal{G}_A$ is performed by an interpolation operator (trilinear \edt{and bilinear $\mbox{3D}$ and $\mbox{2D}$ spaces, respectively}) that is constructed \emph{once}, during initialization, and reused at every transfer thereafter. To keep the interpolation accurate, the acoustic-grid spacing within the overlap region is constrained relative to the local flow spacing\edt{. T}he maximum acoustic cell size in the interpolation zone does not exceed $1.5$~times the local flow cell size. Beyond the interpolation region\edt{,} the acoustic grid is \edt{stretched in a manner that dictates its} coarser resolution purely by acoustic wavelength\edt{-related} requirements\edt{\textit{i.e.,}} 10 points per wavelength. This constraint bounds the ratio of coarse-to-fine spacing seen by the operator and thereby limits the \edt{potential} interpolation error \edt{caused due to mapping the parameters from a} fine flow \edt{grid} onto the acoustic mesh.

Points of $\mathcal{G}_A$ that fall outside $\mathcal{G}_F$ are flagged and assigned zero weight, so that the acoustic field is forced only where valid flow data exist. The truncation of the source at the edge of the overlap region can, if handled naively, inject non-physical energy into the acoustic grid. To suppress \edt{it}, and following the source-filtering treatment of Ewert and
Schr\"oder~\cite{ewert2003acoustic}, the acoustic source is smoothly damped towards zero over a prescribed band of cells near the boundary of the interpolation region using a Gaussian weighting function. The Gaussian taper is found to behave more robustly across a range of cases than a simple linear ramp\edt{. This strategy greatly helps avoid} the spurious reflections that a sharp cut-off would otherwise produce at the edge of the interpolation zone.

\begin{figure*}
    \centering
    \includegraphics[width=1.0\linewidth]{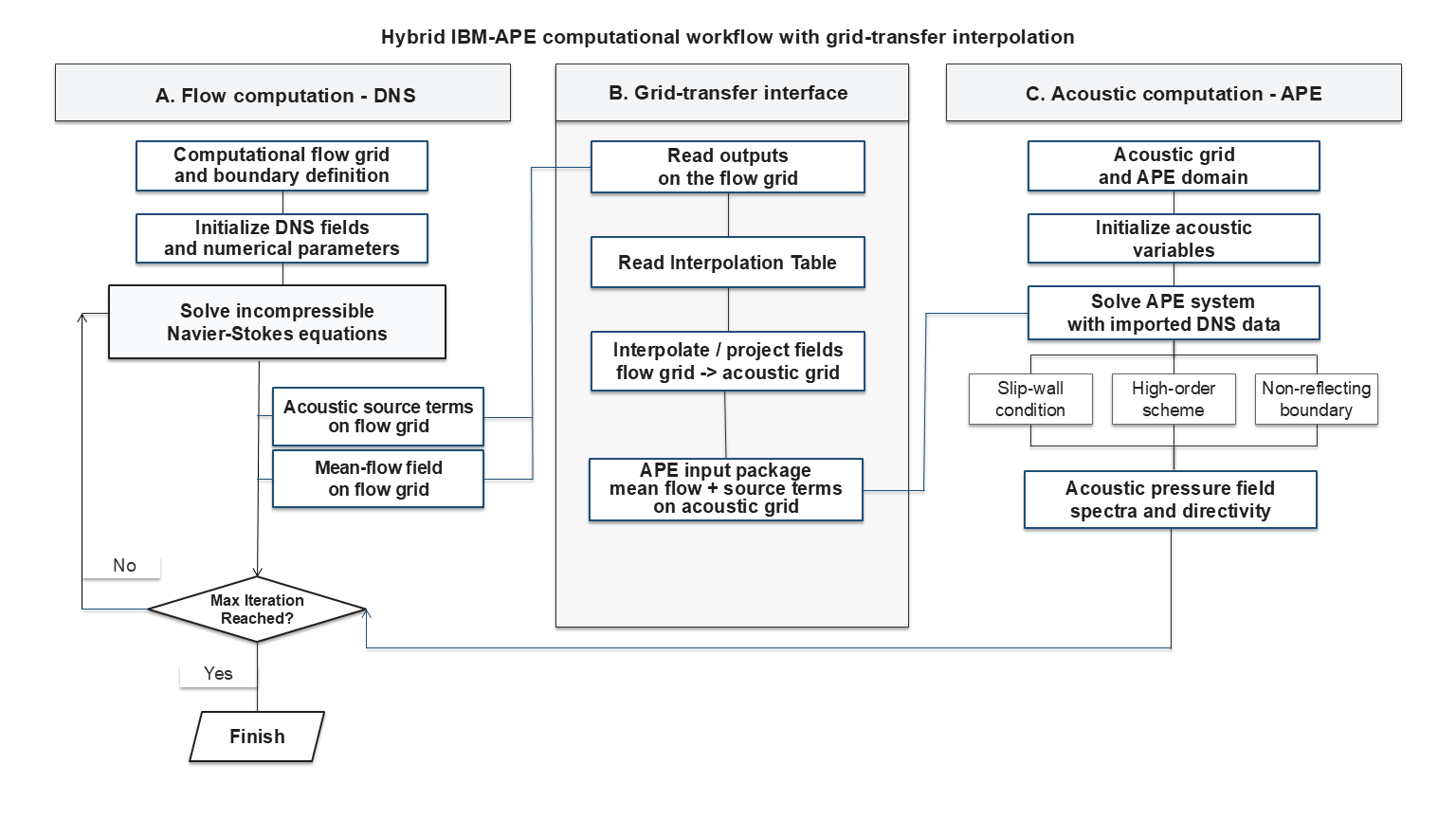}
    \caption{Flowchart of the solution procedure for the IBM-APE dual-grid method.}
    \label{fig:DNS_APE}
\end{figure*}

The distinguishing feature of the present operator is that all of this
geometric \edt{computations, search for cells, evaluation of weights}, and, most importantly, the resolution of cross-process data dependencies, is precomputed and stored in a per-point interpolation table. The eight flow nodes required by an acoustic point need not reside on the rank that owns that point, because the two grids are decomposed independently\edt{, as explained previously}. For every corner\edt{,} the operator determines the owning rank of the flow node from the analytic decomposition rule \edt{(see Eq.}~\ref{eq:decomp}) and classifies it as \emph{local} or \emph{remote}. The remote requests are de-duplicated and aggregated per rank, and a symmetric communication schedule is established once via a single collective exchange of message sizes (\texttt{MPI\_Alltoall}), after which each rank knows exactly which flow nodes it must \edt{be sent} to, and \edt{received} from, every partner. This one-time setup converts an irregular, point-by-point dependency into a fixed, sparse, neighbour-to-neighbour communication pattern. At every transfer \edt{of information,} the operator executes in two inexpensive stages. First, the off-rank flow data are gathered in a single round of non-blocking point-to-point messages (\texttt{MPI\_Isend}/\texttt{MPI\_Irecv} closed by one
\texttt{MPI\_Waitall}), packing only the precise nodes each partner requested
into contiguous buffers and unpacking the received values into a compact
remote-data array. Second, every local acoustic point evaluates its stored
eight-term weighted sum, reading each corner either from local memory or directly from the gathered remote array through a precomputed index. No search, no \edt{recomputation of weights}, and no per-point reduction occur at run time, with the sole exception of cells at the immersed-geometry interface, whose status may switch between solid and fluid as the flow evolves\edt{.} These \edt{interfacial} points are updated online without altering the precomputed tables. \edt{Hence, the computational} cost of a transfer is one sparse, halo-like exchange plus a single pass over the acoustic cells. Because the interpolation table and the  communication schedule depend only on the geometry of the two \edt{fixed} grids, they are built once and amortised over the entire \edt{computations for the solutions}. The complete solution procedure is also summarized in Fig.~\ref{fig:DNS_APE}.

Together, the dual-grid decomposition and the precomputed parallel interpolation operator allow the energetic flow and the weak acoustic field to be resolved\edt{,} each on a grid tailored to its own scales, on independent and well-balanced domain decompositions\edt{. During this whole process,} the data transfer between them, which is normally the bottleneck of overset and multi-grid acoustic couplings, is reduced to a fixed sparse communication and a single arithmetic sweep per transfer.

\section{Results and Discussion}
\label{sec:resrults}

\edt{We validate our multiphysics solver through four physical problems} of increasing complexity. The first case \edt{relates to} the free propagation of a Gaussian pressure pulse in a uniform mean flow and isolates the acoustic discretization together with the far-field non-reflecting treatment, with no immersed body present. The second case \edt{belongs to} scattering of a Gaussian pulse by a rigid circular cylinder, while {adding} the ghost-cell boundary treatment on the acoustic grid and tests it against an analytical solution. The third case \edt{is for} the tonal sound radiated by flow past a circular cylinder, \edt{involving an} incompressible flow, dual-grid source transfer, and acoustic computations, \edt{presenting a coupled problem of fluid-structure-acoustic interactions} and demonstrating \edt{our newly introduced} dual-grid strategy. Finally, the fourth case considers \edt{the flow-induced noise around an undulating foil. \af{This case not only validates the solver for low-Mach-number flows involving undulatory motion, but also demonstrates the effectiveness of the dual-grid strategy through a direct comparison with the single-grid approach.} Moreover, we also present the employment of our newly presented multiphysics solver and its effectiveness for capturing the fluid-structure-acoustic interactions around different marine swimmers, performing specific undulatory kinematics. These species considered here possess morphologies of varying complexity in solitary and schooling configurations. Also, they belong to different classes of marine animals and include \af{(i) a single jack fish (Crevalle Jack),} (ii) an anguilliform swimmer (American eel) in isolation and in four-member school in a diamond configuration, (iii) a manta ray representing the class of swimmers experiencing lift-based propulsion, and (iv) a harbour seal with intricate physiological features as its multiple fore- and hind flippers for producing thrust through intense vortex-body interactions.} \af{The selected swimmers further demonstrate the versatility and effectiveness of the proposed solver in directly capturing their acoustic signatures without relying on simplified acoustic analogies. From a computational perspective, the solver is shown to accommodate a wide range of morphological and kinematic complexities, ranging from the relatively thin (membrane), flexible body of the manta ray to the fully volumetric and physiologically complex body of the harbor seal, and from the predominantly tail-driven propulsion of the jack fish to the full-body undulatory motion of the eel in both solitary and schooling configurations. This diversity of test cases demonstrates the robustness of the solver in handling markedly different fluid-structure-acoustic interaction scenarios, while also providing detailed information on the acoustic signatures and spatial footprints associated with each type of marine swimmer.}



\subsection{Propagation of a Gaussian acoustic pulse}
\label{sec:val_pulse}

\begin{figure}
    \centering
    \includegraphics[width=0.6\linewidth]{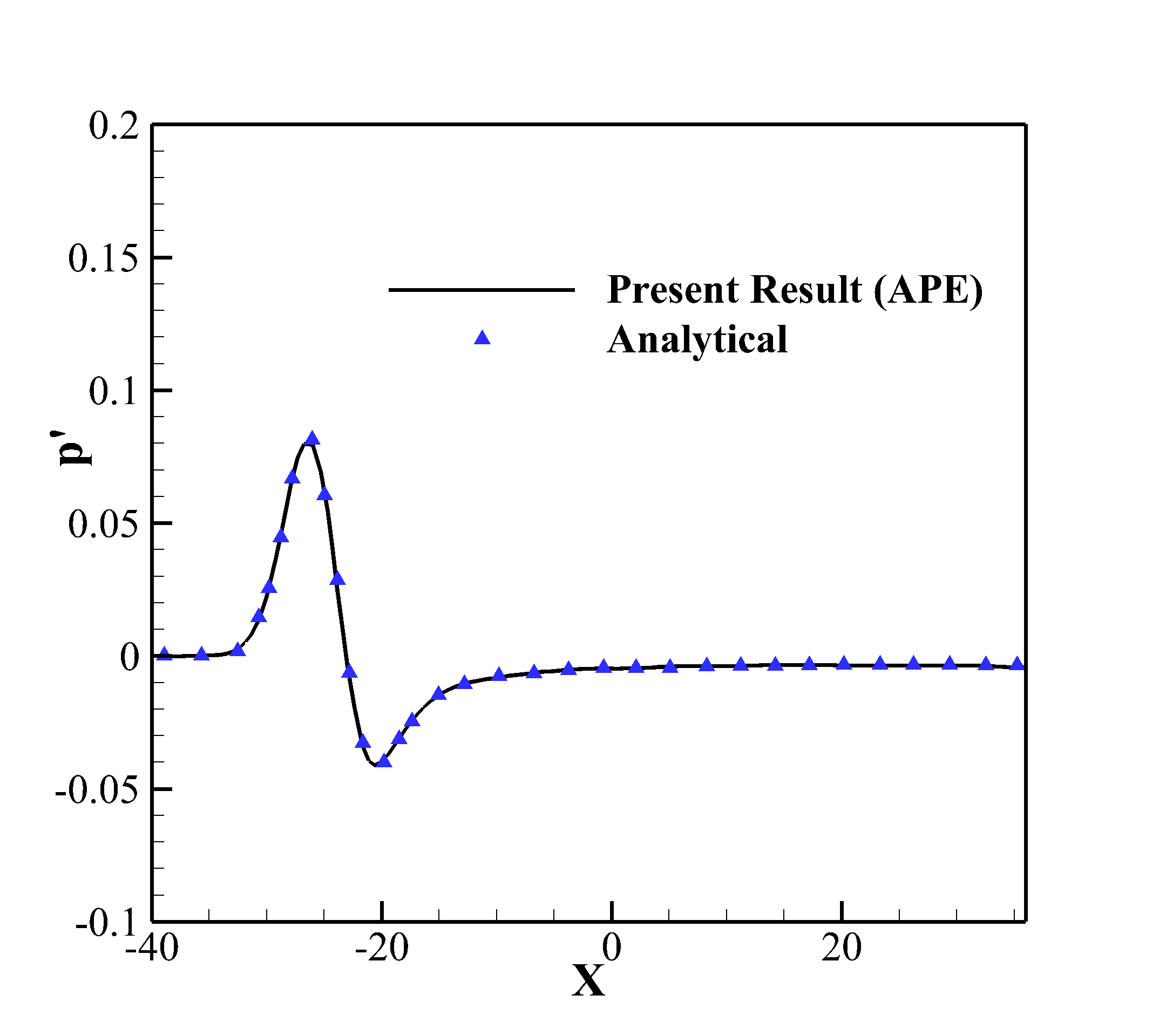}
    \caption{Acoustic pressure fluctuations along the x-axis at $t=50 \mbox{s}$}
    \label{fig:Val_Pure_Pulse}
\end{figure}

The first case examines the free propagation of a two-dimensional Gaussian
pressure pulse \cite{hardin1994icase} superposed on a uniform mean flow at $\mathrm{Ma}=0.5$. Because no solid body is present \edt{within the flow field}, the \edt{scenario} isolates the accuracy of the DRP spatial operator, the LDDRK\edt{-based} time integration, and the ETA non-reflecting boundary treatment. \edt{With a half-width of $b = 3$,} the pulse is initialized at the centre of a square domain $-50 \le (x,y) \le 50$ \edt{with the following specifications:}
\begin{equation}
p' = \exp\!\left[-\ln 2 \,\frac{x^2 + y^2}{b^2}\right], \qquad
u' = v' = 0
\label{eqn:val_pulse_init}
\end{equation}
\af{An exact solution for the subsequent evolution is provided by Hardin et al. \cite{hardin1994icase} (Category~1, Problem~1) and is adopted here as the reference solution.} 


As the pulse \edt{grows,} it spreads isotropically at the local wave speed, \edt{which is} the sum of the \edt{speed of sound} and the mean-flow velocity, so that the wavefront is convected downstream while remaining circular, and passes cleanly through the outer boundary with no visually discernible reflection. As evident in Fig.~\ref{fig:Val_Pure_Pulse}, the pressure profile along \edt{the axis of} $y=0$ at \af{$t=50~\mathrm{s}$}  from the APE \edt{exactly matches with the} analytical \edt{solution profile}. \edt{This comparison confirms} that the propagation model, the high-order scheme, and the radiation boundary condition are consistent with one another.

\subsection{Scattering of a Gaussian pulse by a circular cylinder}
\label{sec:val_scatter}

\edt{Now,} the second case introduces an immersed body \edt{in an acoustic field} to test the \edt{presently proposed} ghost-cell acoustic boundary treatment. A rigid circular cylinder of diameter $D = 1$ is centered at the origin, and a Gaussian pulse \edt{with the following parameters} is released from $(x,y) = (4,0)$ \af{\cite{tam1997second}}\edt{:}

\begin{equation}
p' = \exp\!\left[-\ln 2 \,\frac{(x-4)^2 + y^2}{0.2^2}\right], \qquad
u' = v' = 0
\label{eqn:val_scatter_init}
\end{equation}
\noindent within a domain $-6 \le (x,y) \le 6$ discretized by a uniform mesh $\Delta x = \Delta y = 0.02$. On the cylinder\edt{'s} surface\edt{,} the slip condition $\mathbf{u}^{a}\!\cdot\!\mathbf{n}=0$ and the homogeneous Neumann condition $\partial p'/\partial n = 0$ are imposed through the ghost cells. The pressure \edt{profile} is monitored at two \edt{locations} ($A(2,0)$, $B(2,2)$) for comparison with the analytical solution \edt{of this problem}. \af{An analytical solution is provided by Tam and Hardin \cite{tam1997second} (Category~1, Problem~2) and is adopted here as the reference solution.}


\begin{figure*}[t]
    \centering

    \begin{subfigure}[b]{0.32\textwidth}
        \centering
        \includegraphics[width=\linewidth]{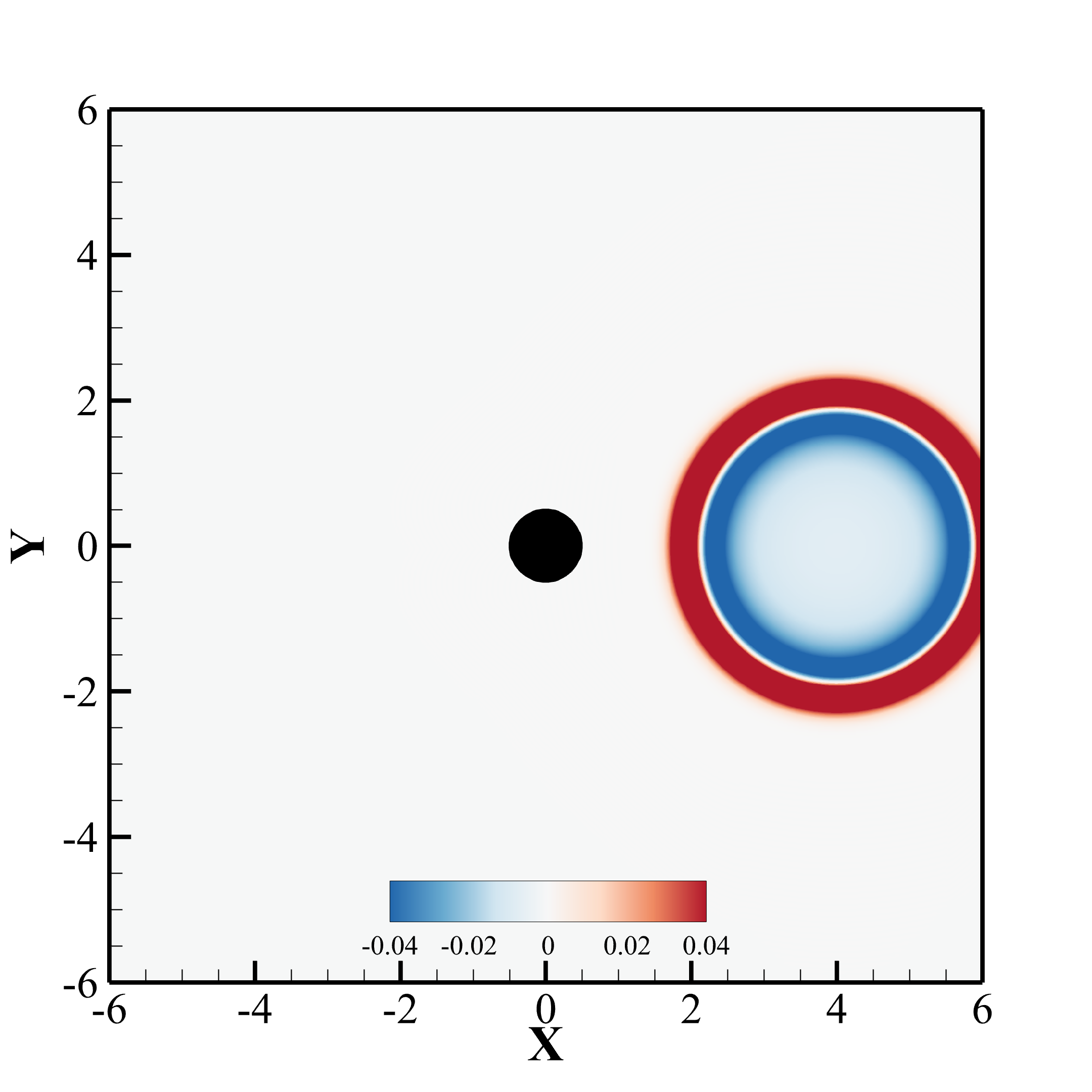}
        \caption{}
        \label{fig:pulse_t2}
    \end{subfigure}
    \hfill
    \begin{subfigure}[b]{0.32\textwidth}
        \centering
        \includegraphics[width=\linewidth]{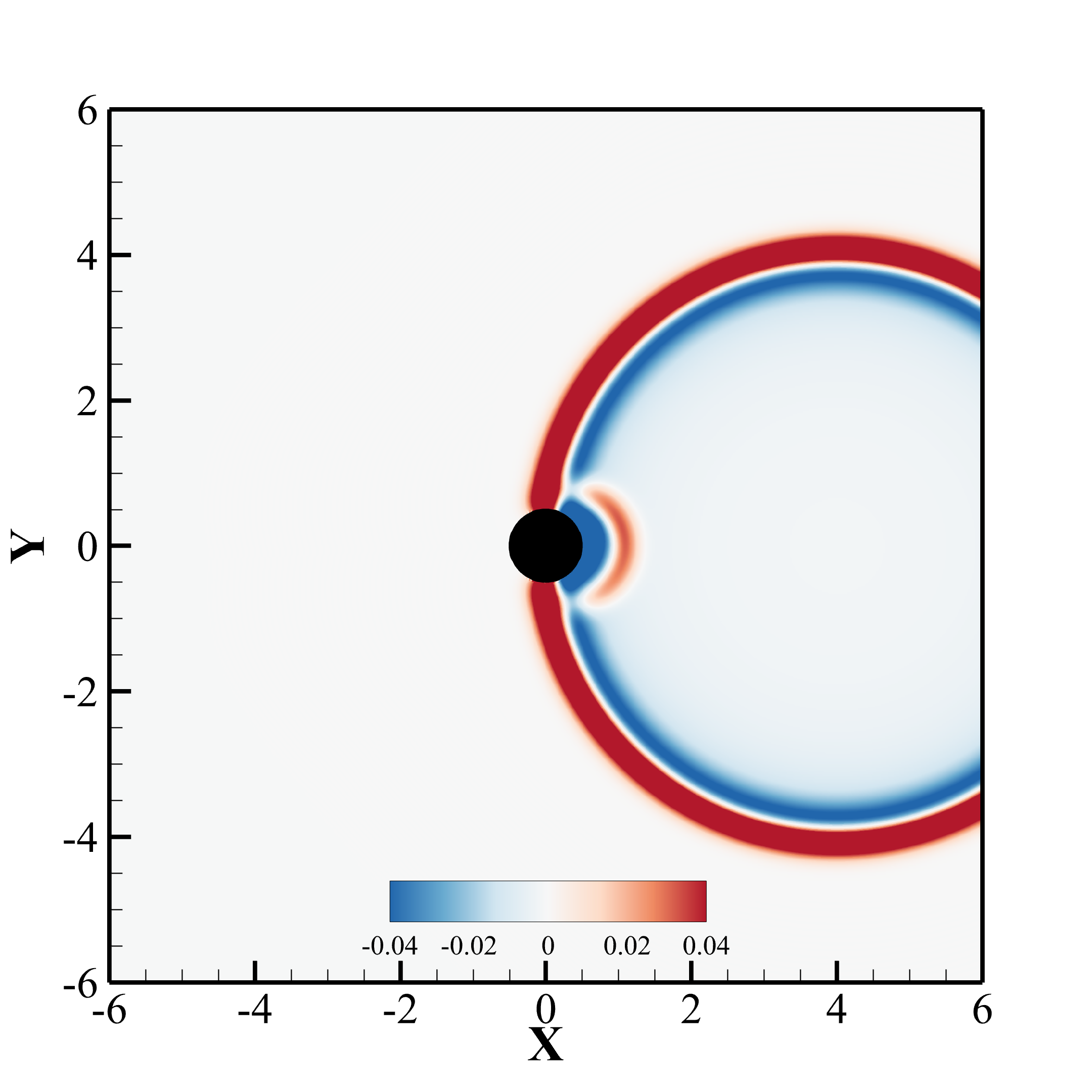}
        \caption{}
        \label{fig:pulse_t4}
    \end{subfigure}
    \hfill
    \begin{subfigure}[b]{0.32\textwidth}
        \centering
        \includegraphics[width=\linewidth]{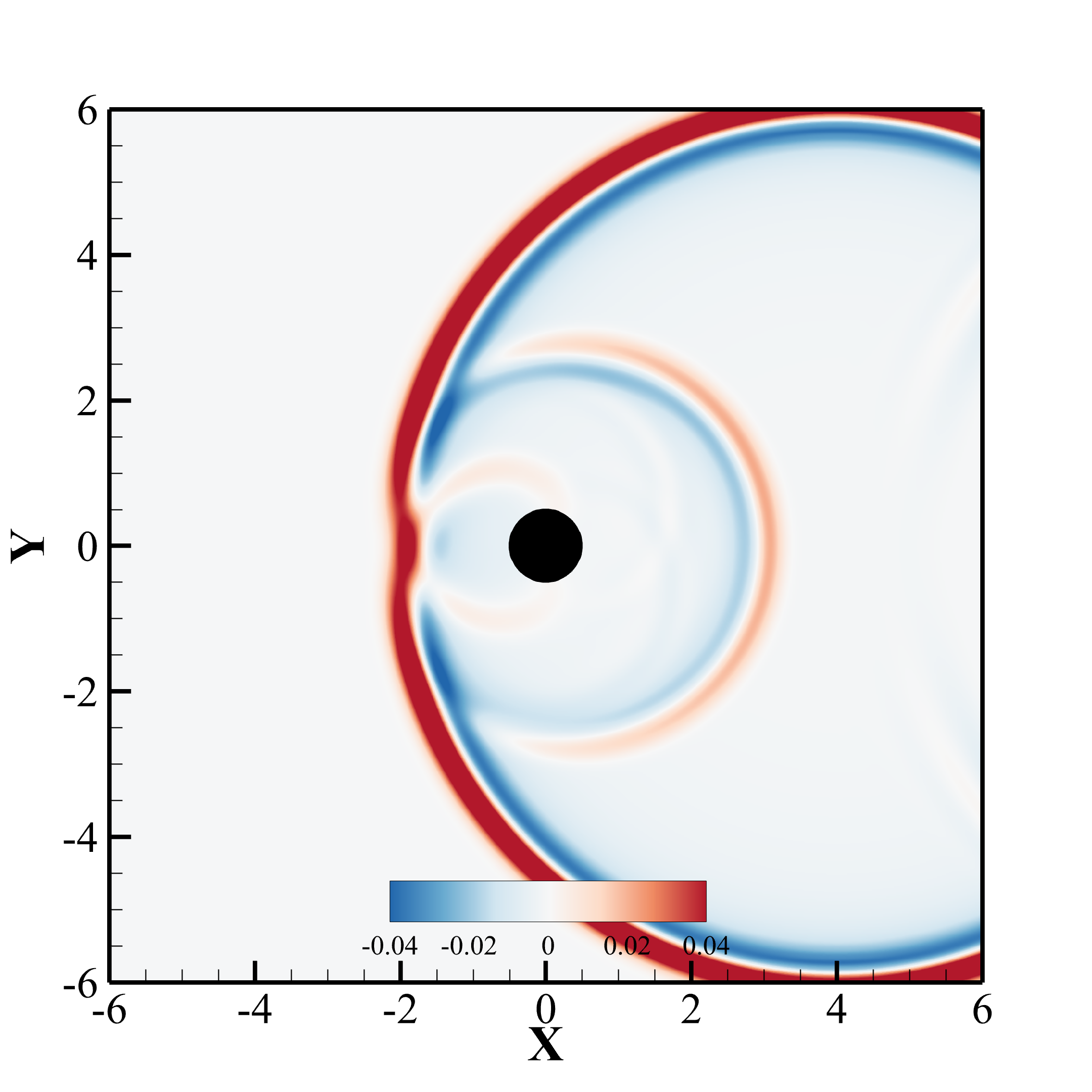}
        \caption{}
        \label{fig:pulse_t6}
    \end{subfigure}

    \vspace{0.0cm}

    \begin{subfigure}[b]{0.7\textwidth}
        \centering
        \includegraphics[width=\linewidth]{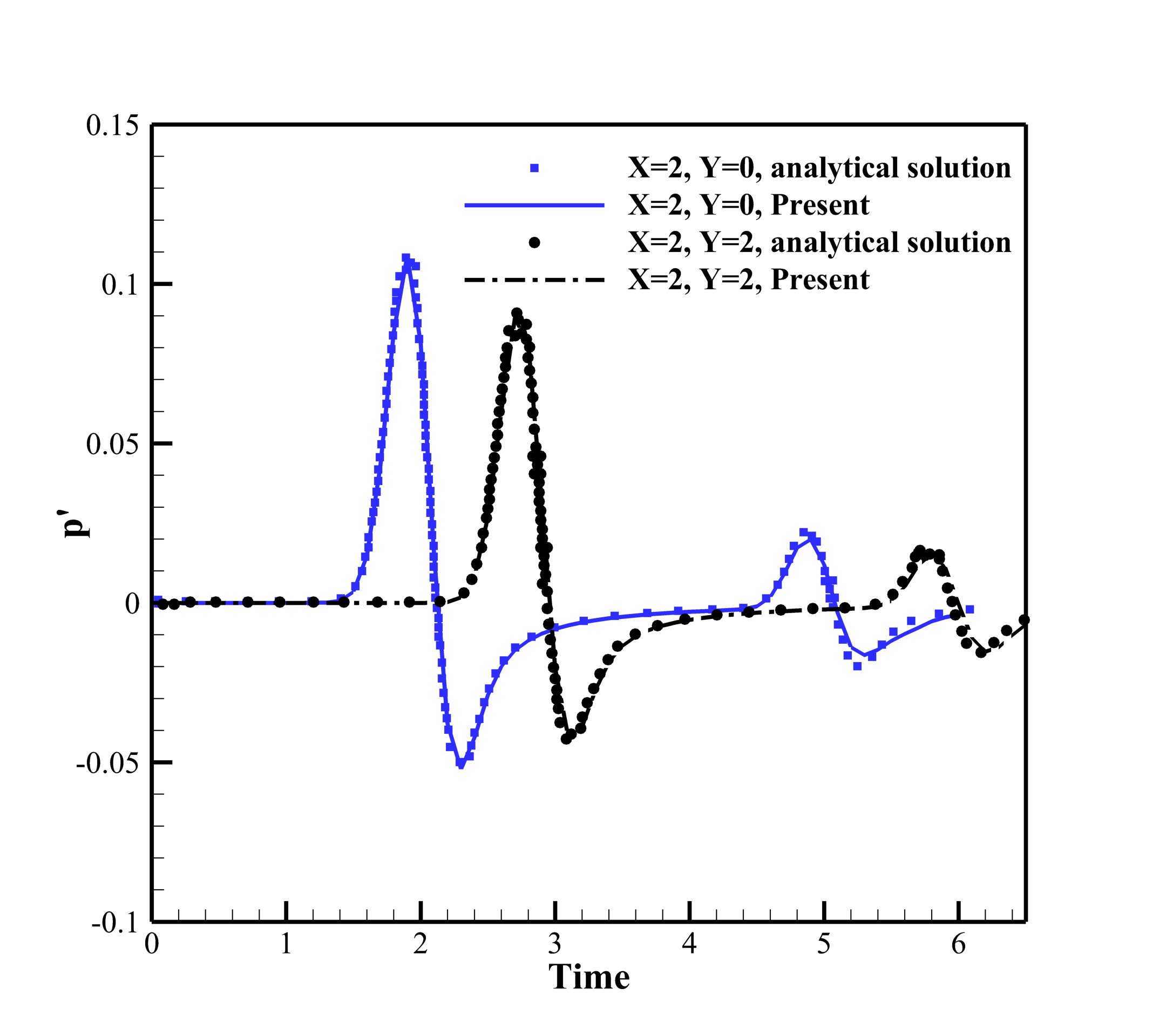}
        \caption{}
        \label{fig:pulse_comparison}
    \end{subfigure}

    \caption{{Acoustic pressure fluctuation contours} at
    (a) $t=2$, (b) $t=4$, and (c) $t=6$.
    (d) Comparison between the present and analytical solutions from
    $t=0$ to $t=6$ at observation points A $(2,0)$ and B $(2,2)$.}
    
    \label{fig:Val_Cyl_Pulse}
\end{figure*}

The incident pulse strikes the cylinder, \edt{gets reflected} from the right side of the \edt{cylinder} back towards the source, and diffracts around the \edt{farther part of the cylinder}, producing the \edt{evolving} multi-lobed patterns \edt{shown} in \edt{Figs.~\ref{fig:Val_Cyl_Pulse}a, \ref{fig:Val_Cyl_Pulse}b, and \ref{fig:Val_Cyl_Pulse}c}. \edt{The comparative plots in Fig.~\ref{fig:Val_Cyl_Pulse}d exhibits that the} \edt{computed temporal profiles of the fluctuating acoustic pressure recorded} at the two \edt{identified locations} produce both the incident and the reflected wave, and agree \edt{very well} with \edt{those from the} analytical solution in \edt{terms of both} amplitude and phase. \edt{These results demonstrate} that the ghost cell\edt{-based} reconstruction imposes the rigid-wall condition accurately on the non-conforming Cartesian mesh.

\subsection{Tonal sound generated by flow past a circular cylinder}
\label{sec:val_aeolian}

The third case \edt{explains the flow-induced noise in the presence of a bluff body in the flow field. Dealing with this case requires activating computations of the incompressible flow field, the one-wat dual-grid strategy for transfering the source to the acoustic field} and the acoustics\edt{-related computations} on a separate, larger grid. The Reynolds number based on the cylinder\edt{'s} diameter is $200$, and the Mach number is $0.2$\edt{. T}his configuration \edt{represents} a standard benchmark \edt{problem} for hybrid aeroacoustic solvers \cite{seo2011high,zhao2021sharp}. The incompressible flow is solved first with the sharp-interface IBM solver\edt{. When the solution} reaches \edt{its} periodic \edt{steady} state, the time-averaged mean field and the acoustic source term $\mathbf{S}_{\mathrm{mom}} = \nabla(P-\bar{P})/\rho_0$ are transferred to the acoustic grid and the APE system is \edt{solved}.

\begin{figure*}
    \centering
    \includegraphics[width=1.0\linewidth]{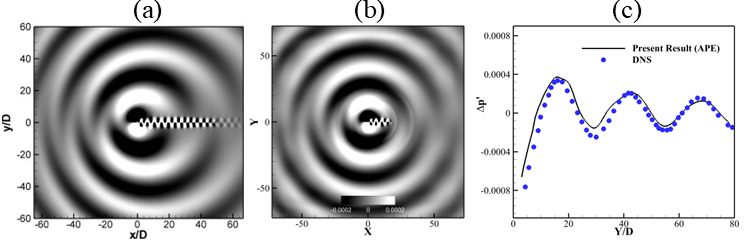}
    \caption{Comparison of instantaneous acoustic pressure fluctuation contours, non-dimensionalized by $\rho_0 c^2$. (a) direct simulation of the compressible Navier--Stokes equations on the body-fitted O-type grid \af{\cite{seo2011high}} \edt{,} (b) Incompressible Navier-Stokes/APE hybrid method on the Cartesian grid \edt{from the currently presented solver, and} (c) \edt{a comparison of} pressure \edt{fluctuations} $\Delta p'$ along the $x=0$ line above the cylinder, where $\Delta p' = p-\bar{p}=(P+p')-\overline{(P+p')}$}
    \label{fig:Val_Flow_around}
\end{figure*}

\edt{We specifically choose and present this well-studied problem to put our} dual-grid strategy to work. The flow grid is restricted to the region that must resolve the near wake, with a fine near-body spacing of $\Delta x_{\min} = \Delta y_{\min} = 0.02D$\edt{. T}he acoustic grid shares this minimum spacing near the cylinder, so that the source is transferred without \edt{any} interpolation error\edt{,} where accuracy matters most\edt{,} and \edt{our approach captures} the curvature \edt{of the surface well}. \edt{Beyond this region, the acoustic grid is stretched} outward to an acoustic resolution of roughly $10$ points per wavelength, extending to a far-field radius large enough to capture the directivity while \edt{simultaneously} damping the outgoing waves in the ETA layer. Because the two grids are decoupled, this \edt{size of the} far field (\edt{$200D \times 200D$ in this case}) is obtained without enlarging the flow domain (\edt{$40D \times 30D$}). \edt{This exercise effectively demonstrates how our present solver is effective and efficient in terms of reducing the computational cost for solving such multiphysics problems.}

\edt{As presented in Fig.~\ref{fig:Val_Flow_around}, t}he flow \edt{solver} produces the expected von K\'arm\'an vortex street in the wake. In the acoustic field, the periodic lift fluctuations produce the characteristic lift-dipole radiation pattern \edt{that is} oriented normal to the flow (\edt{see Figs.~\ref{fig:Val_Flow_around}a and \ref{fig:Val_Flow_around}b}). \edt{Here, the wavelength of $27D$ for the radiated sound} measured along the transverse direction is consistent with \edt{its} theoretical estimate \edt{of} $\lambda  \approx 26.7D$. Finally, we compare the acoustic pressure \edt{fluctuations} sampled along the transverse centre-line above the cylinder with the reference results \edt{reported by} Seo and Mittal \cite{seo2011high} and Zhao et al. \cite{zhao2021sharp} \edt{in} Fig.~\ref{fig:Val_Flow_around}\edt{c. The} close agreement \edt{of the presently obtained profiles with those in the literature} in \edt{terms of} both amplitude and decay confirms that \edt{our} dual-grid coupling \edt{technique} transfers the source \edt{accurately} and that the complete hybrid solver predicts flow-induced sound accurately.

\subsection{Sound generated by a traveling wavy foil}
\label{sec:val_wavy_foil}

\begin{figure}
    \centering
    \includegraphics[width=1.0\linewidth]{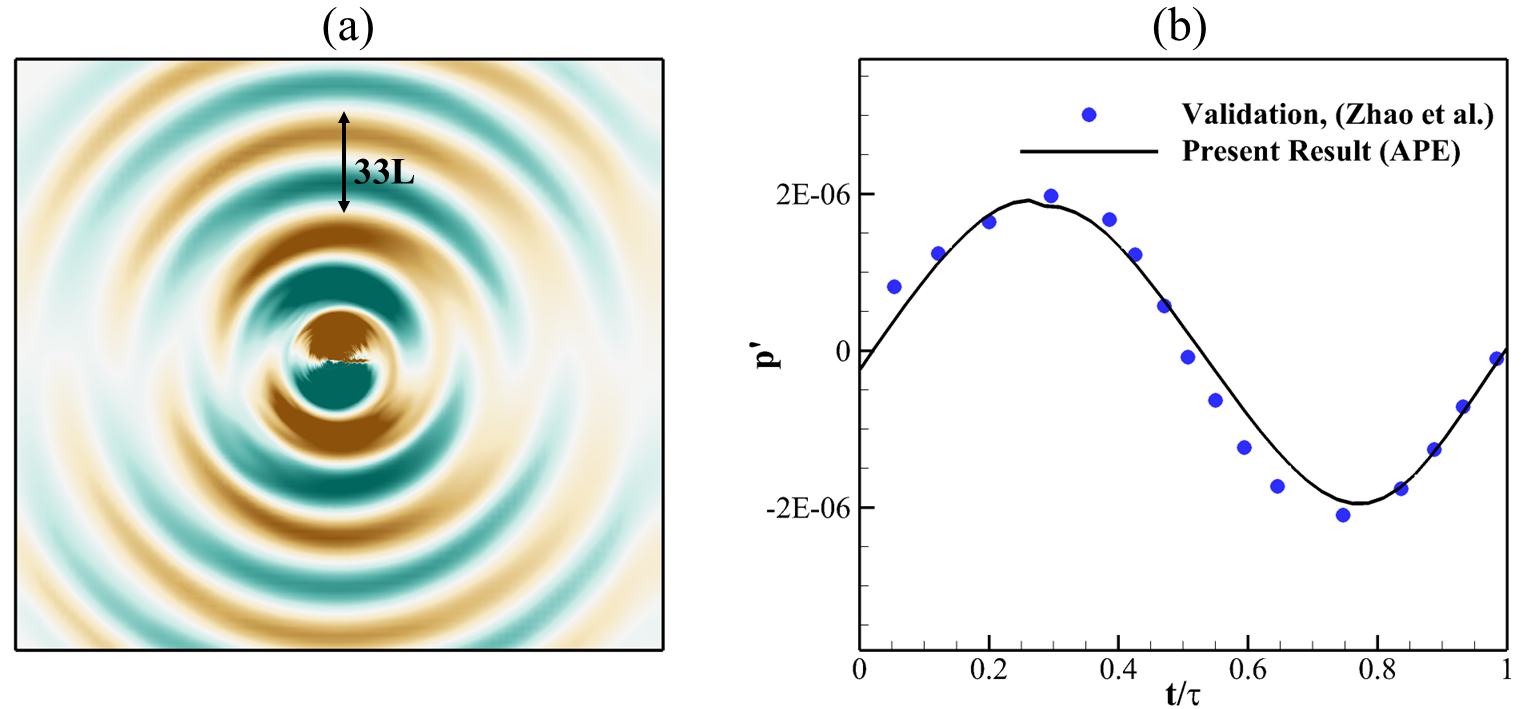}
    \caption{Validation for the LA--HF traveling wavy foil at $St=0.6$ and $\mathrm{Ma}=0.01$. (a) Instantaneous non-dimensional acoustic pressure fluctuation contours; the arrow marks the measured vertical wavelength of approximately $33L$. (b) Acoustic pressure fluctuation at $r=60L$ and $\theta=90^{\circ}$ over one undulation period, comparing the present APE solution with the data of Zhao et al.~\cite{zhao2022sound}. Here, $p'$ is non-dimensionalized by $\rho_0c_0^2$ and $\tau=1/f_0$.}
    \label{fig:Val_Fish}
\end{figure}

The fourth case examines the sound generated by a single traveling wavy foil and validates the treatment of a moving and continuously deforming immersed boundary as we as low mach number. The low-amplitude-high-requency (LA-HF) configuration of Zhao et al.~\cite{zhao2022sound} is reproduced. The foil has a NACA~0012 profile of chord length $L$, and its prescribed lateral displacement is
\begin{equation}
\begin{aligned}
\frac{y(x,t)}{L}
&= a(x)\sin\!\left(2\pi x-2\pi f_0t+\phi\right),\\
a(x)&=0.02-0.08x+0.16x^2,\qquad 0\leq x\leq 1,
\end{aligned}
\label{eqn:val_wavy_kinematics}
\end{equation}
for which the maximum tail amplitude is $a_{\max}=0.1L$. The flow and acoustic parameters are $Re=1000$, the Strouhal number $St=2a_{\max}f_0/U_0=0.6$, and $\mathrm{Ma}=0.01$. As in the preceding coupled validation, the incompressible flow is first advanced to a periodic state, after which the pressure-fluctuation source is supplied to the APE solver. The acoustic pressure is non-dimensionalized by $\rho_0c_0^2$.

The instantaneous pressure field in Fig.~\ref{fig:Val_Fish}(a) displays the vertically oriented dipole pattern reported in Zhao et al.~\cite{zhao2022sound}. For the present kinematics, $St=0.6$ and $a_{\max}/L=0.1$ give $f_0L/U_0=3$; therefore, the acoustic wavelength predicted from the undulatory frequency is
\begin{equation}
\frac{\lambda}{L}
=\frac{c_0}{f_0L}
=\frac{1}{\mathrm{Ma}\,(f_0L/U_0)}
=33.3
\label{eqn:val_wavy_wavelength}
\end{equation}
The wavelength measured from the present pressure contours is approximately $33L$, in close agreement with both this estimate and the reference result. The low Mach number also makes the convective correction to the radiated wavelength negligible.

For a quantitative comparison, the pressure fluctuation is sampled in the vertical direction at $r=60L$ and $\theta=90^{\circ}$. Figure~\ref{fig:Val_Fish}(b) compares the phase-folded signal over one undulation period, $\tau=1/f_0$, with the reference data. The present APE solution reproduces the nearly sinusoidal waveform, including the phase of the zero crossings and the positive and negative extrema. Small pointwise differences occur near the extrema, but the overall amplitude and phase agree closely with the reference. Together with the recovered dipole structure and wavelength, this comparison confirms that the moving-boundary treatment and the coupled flow-acoustic procedure accurately predict the sound radiated by an undulating foil.

\af{To assess the computational overhead associated with the dual-grid approach, the present undulating-foil case was also used to compare the computational cost of the single-grid and dual-grid strategies. This case involved with a low-Mach number flow is particularly suitable for such a comparison because the characteristic spatial requirements of the hydrodynamic and acoustic fields become increasingly challenging to handle as the Mach number decreases. The flow-related solution requires sufficient resolution primarily in the vicinity of the body and its wake, whereas the acoustic grid must extend over a substantially larger region to accurately capture the generation and propagation of acoustic waves. Consequently, the use of a single grid requires the flow equations to be solved over a domain whose extent is dictated partly by the acoustic propagation requirements, resulting in unnecessary computational effort.

For a consistent comparison, both simulations were performed using the same $4\times2$ domain decompositions. In the single-grid configuration, a mesh with $641\times641$ in the $x$- and $y$-axes was used for both the flow and acoustic solvers. In the dual-grid configuration, the flow equations were solved on a smaller mesh of size $561\times561$, while the acoustic equations were solved on the original mesh with $641\times641$ nodes. The acoustic source terms were transferred from the flow grid to the acoustic grid through the interpolation procedure described previously. The corresponding computational times per iteration are summarized in Table~\ref{tab:computational_cost}. Reducing the size of the flow grid from $641\times641$ nodes to $561\times561$ nodes results in a corresponding reduction in the computational time of the flow solver from approximately $0.064$~s to $0.050$~s per iteration. On the contrary, the computational time of the acoustic solver remains essentially unchanged because the same acoustic grid is employed in both configurations. More importantly, the additional interpolation required by the dual-grid approach requires only approximately $0.00109$~s per iteration, which is negligible compared with the computational costs of both the flow and acoustic solvers.

These results demonstrate that the interpolation required by the dual-grid formulation introduces only a minor computational overhead while allowing the flow and acoustic grids to be designed independently according to their respective physical requirements. Particularly, the flow grid can be concentrated around the body and wake, where the relevant hydrodynamic structures must be accurately resolved, whereas the acoustic grid can be independently designed to resolve the acoustic generation and propagation regions. This flexibility becomes particularly advantageous for low-Mach-number aero- and hydro-acoustic problems, for which the spatial requirements of the hydrodynamic and acoustic fields can differ substantially. }

\begin{table}[t]
    \centering
    \caption{Comparison of the computational cost of the single-grid and dual-grid strategies for the undulating-foil case. The reported values represent the computational time per iteration using the same $4\times2$ domain decomposition.}
    \label{tab:computational_cost}
    \begin{tabular}{lccc}
        \hline
        Strategy & Flow solver (s) & Acoustic solver (s) & Interpolation (s) \\
        \hline
        Single-grid & 0.064 & 0.340 & -- \\
        Dual-grid   & 0.050 & 0.350 & 0.00109 \\
        \hline
    \end{tabular}
\end{table}

\subsection{Single eel and Jack fish}


\label{sec:case_eel_jack}

Following the four validation cases, we consider four biological biological aquatic swimmers, belonging to different classes and serving as representatives of entirely different morphologies and undulatory kinematics.
{Operating at low Mach numbers, different swimming modes produce substantially different body kinematics and wake structures. Carangiform swimmers, such as jack fish considered here, generate thrust predominantly through large-amplitude oscillations of the posterior body and caudal fin, whereas the anterior portion of the body undergoes relatively small lateral motion \cite{khalid2021larger}. In contrast, anguilliform swimmers, such as eels exhibit pronounced undulatory motion over nearly the entire elongated body, resulting in a fundamentally different fluid-structure interaction and wake topology \cite{khalid2021anguilliform}. Therefore, these two swimmers  provide complementary configurations for evaluating the newly proposed dual-grid framework over a broad range of bio-inspired swimming kinematics and morphologies. The flow grid is designed to provide sufficient spatial resolution near the moving body and in the hydrodynamic wake, while the independently constructed acoustic grid enables the weak acoustic disturbances generated by the swimmer to be resolved and propagated efficiently over a considerably larger domain. Here,} each swimmer has a characteristic length $L$ and is placed at the center of the computational domain. Following Khalid et al.~\cite{khalid2021anguilliform,khalid2021larger}, the eel's amplitude envelope is

\begin{equation}
 A_{\mathrm{eel}}(x)
 =0.0367+0.0323x+0.0310x^2
 \label{eq:eel_envelope}
\end{equation}

\noindent whereas the jack fish uses the carangiform motion envelope:

\begin{equation}
 A_{\mathrm{Jack}}(x)
 =0.02-0.0825x+0.1625x^2.
 \label{eq:jack_envelope}
\end{equation}

For both swimmers, the prescribed lateral displacement has the traveling-wave
form

\begin{equation}
 z(x,t)=A(x)\sin\!\left[2\pi
 \left(\frac{x}{\lambda}-ft\right)+\phi\right].
 \label{eq:eel_jack_motion}
\end{equation}

The nondimensional wavelengths are $\lambda^*_{\mathrm{eel}}=0.95$ and
$\lambda^*_{\mathrm{Jack}}=1.05$, and both cases use $St=0.4$,
$\mbox{Re}=3000$, and $\mathrm{Ma}=0.1$. The flow domain is a cube with each side side $10L$ long with $426\times354\times354$ grid points. The acoustic grid contains $490\times418\times418$ nodes and extends over a cube with each side of length $80L$. Far-field pressure is sampled in the $x$$y$-plane at $r=15L$.

\begin{figure*}[t]
 \centering
 \includegraphics[width=\textwidth]{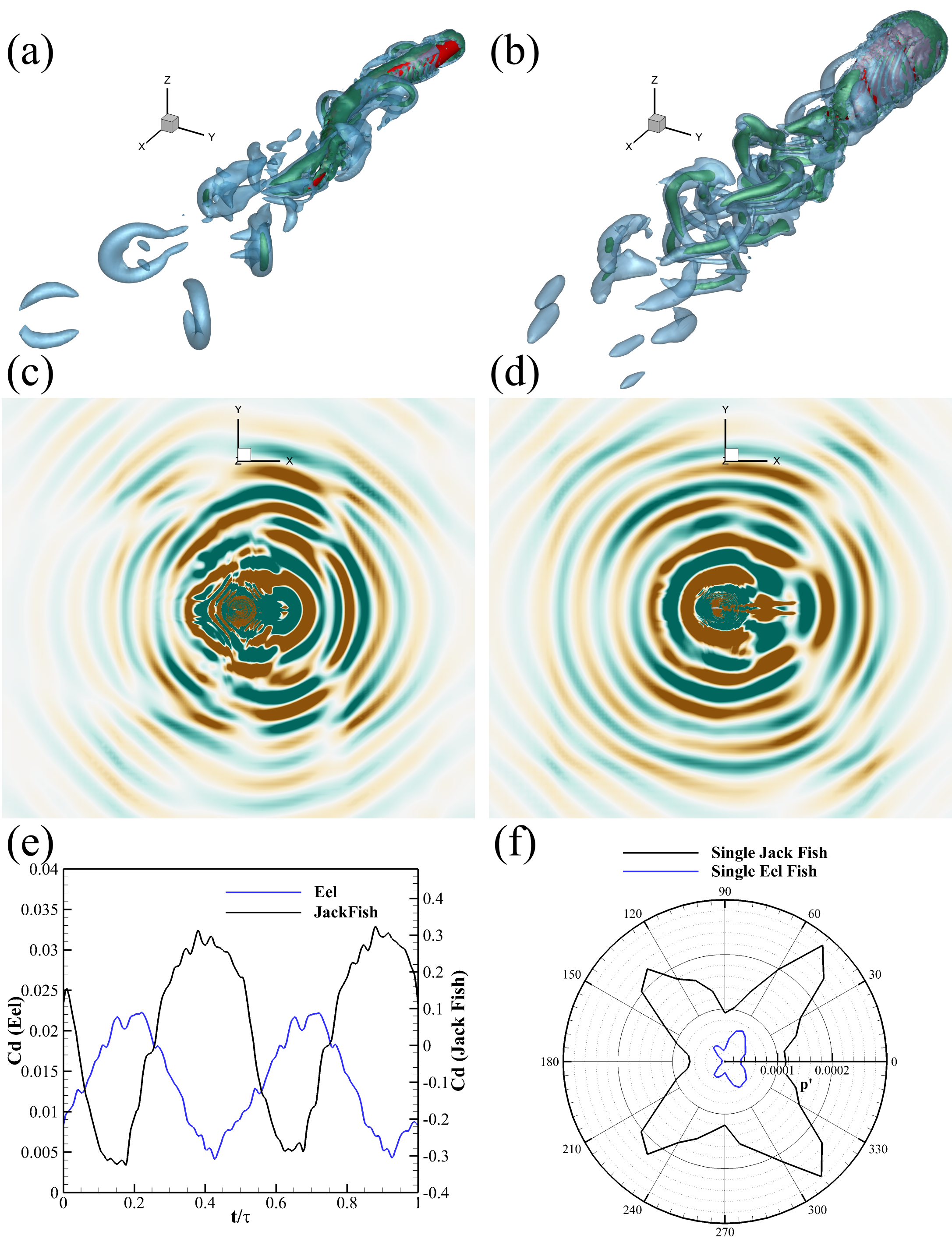}
 \caption{Comparison between flow, acoustics, and forces produced by a single eel and a single Jack fish, where instantaneous $Q$-criterion structures for (a) the eel and (b) the jack fish, are shown by $Q=10$ in blue and $Q=60$ in green. Instantaneous nondimensional acoustic pressure on the $xy$-plane for (c) the eel and (d) the jack fish, (e) drag-coefficient histories  over one undulation period, where the left and right axes correspond to the eel and jack fish, respectively, (f) directivity and magnitudes of far-field acoustic-pressure directivity at $r=15L$.}
 \label{fig:Single_Eel_Jack}
\end{figure*}

Figures~\ref{fig:Single_Eel_Jack}a and \ref{fig:Single_Eel_Jack}b show that the solver resolves the different three-dimensional wakes produced by the eel's slender body and the Jack fish's trunk and fins. The associated acoustic pressure fields, shown in Figs.~\ref{fig:Single_Eel_Jack}c and \ref{fig:Single_Eel_Jack}d) contain well-resolved outgoing wavefronts across the enlarged acoustic domain. The drag histories are periodic for both swimmers, although their magnitudes and phases differ and are therefore displayed with separate vertical scales (Fig.~\ref{fig:Single_Eel_Jack}e). Most notably, the jack fish produces a substantially larger far-field pressure amplitude than the eel at the same location. Both acoustic fields are directional, with the Jack showing four prominent oblique lobes (Fig.~\ref{fig:Single_Eel_Jack}f). These differences demonstrate that the coupled formulation retains the effects of body morphology and prescribed gait from the near-body flow solution through to the acoustic far field. \edt{It is noteworthy here this acoustic spectrum is different from the ones presented by Zhou et al. \cite{zhou2024effect}, where their results show an weak dipole (more like an elongated monopole-like) acoustic spectrum for a single carangiform swimmer (mackerel in their case), and our results show more detailed biased quadropole-like acoustic behavior.}

\subsection{A four-member school of eels in diamond formation}
\label{sec:case_school}

The second application places four identical eels in a diamond formation and compares their collective radiation with that of the single eel.
{This configuration is selected to represent a more realistic biological scenario, since many fish species exhibit schooling behavior in their natural habitats \cite{ligman2024comprehensive}. Compared with an isolated swimmer, a group of swimmers introduces additional hydrodynamic and acoustic interactions, as the wake and pressure disturbances generated by each individual can interact with those produced by neighboring swimmers. From the computational perspective, this case also demonstrates the flexibility of the proposed coupled flow--acoustic framework. Within the APE-based acoustic solver, the sound generated by all four swimmers, together with the resulting interference and propagation of their acoustic disturbances, can be resolved simultaneously without requiring any additional acoustic modeling procedure beyond that used for the single-swimmer case. Hence, the extension from a single eel to a school of multiple swimmers primarily involves the introduction of additional moving geometries, while the same numerical framework is retained for evaluating the collective acoustic radiation.}

The body's model, kinematics, $\lambda^*=0.95$, $\mbox{St}=0.4$, $\mbox{Re}=3000$, and $\mathrm{Ma}=0.1$ are unchanged from Section~\ref{sec:case_eel_jack}. The same flow domain and $80L$ acoustic domain, with $426\times354\times354$ nodes and $490\times418\times418$ grid points, respectively, are retained. Therefore, this case tests the capability of our solver to handle multiple immersed bodies without changing the numerical formulation or far-field treatment.

\begin{figure*}[t]
 \centering
 \includegraphics[width=\textwidth]{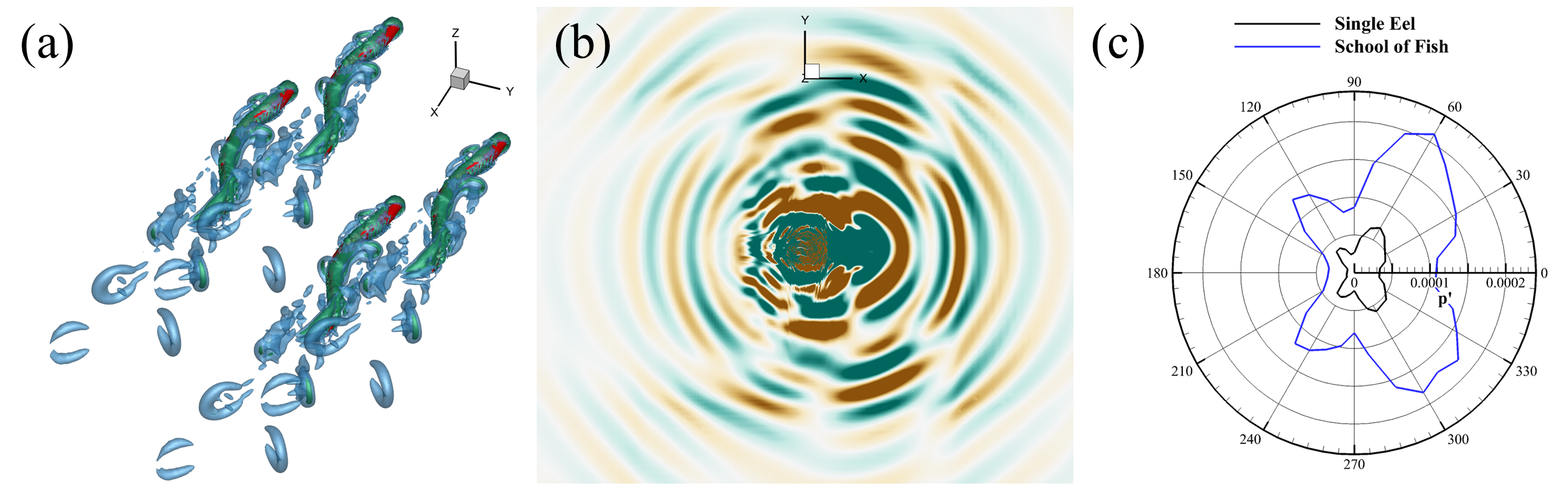}
 \caption{Four eels in a diamond formation: (a) instantaneous $Q$-criterion
 structures, with $Q=10$ in blue and $Q=60$ in green; (b) instantaneous
 nondimensional acoustic pressure on the $x$--$y$ plane; and (c) far-field
 acoustic-pressure directivity at $r=15L$, compared with the single-eel case.}
 \label{fig:School}
\end{figure*}

The wake visualization in Fig.~\ref{fig:School}a shows that coherent structures are captured around each swimmer and remain resolved as the individual wakes interact. Their superposed acoustic field forms a more spatially complex pattern than that of an isolated eel, with clear constructive and destructive interference between waves emitted by different members, as presented in Fig.~\ref{fig:School}b. Consistent with this superposition, Fig.~\ref{fig:School}c exhibits that the school has a larger pressure amplitude over most observation angles and a more pronounced, multi-lobed directivity than the single eel. The result demonstrates that the solver can handle several moving bodies and carry their combined unsteady source field to the acoustic grid.

\subsection{Manta ray}
\label{sec:case_manta}

The third application considers a manta ray, representing a fundamentally different mode of aquatic propulsion from the body-caudal fin swimming mechanisms examined in the preceding cases. Manta rays employ lift-based propulsion, in which thrust is generated primarily through traveling-wave deformations and oscillatory motion of their broad pectoral fins. This case is particularly important because it introduces substantially more complex three-dimensional kinematics and a thin, flexible, membrane-like morphological structure, providing a distinct challenge for the numerical treatment of moving immersed boundaries. Here, we prescribe its prescribed motion following Zhang et al.~\cite{zhang2022vortex}. If $(x_f,y_f,z_f)$ denotes a point on the neutral, flat configuration of the fins, its deformed position is:

\begin{equation}
\begin{aligned}
 x(x_f,y_f,t)&=x_f,\\
 y(x_f,y_f,t)&=y_f\left(1-(1-k)
 \frac{|g(x_f,t)|y_f}{SL}\right)
 \cos\!\left(\frac{\theta_{\max}y_f}{SL}g(x_f,t)\right),\\
 z(x_f,y_f,t)&=z_f+y_f\left(1-(1-k)
 \frac{|g(x_f,t)|y_f}{SL}\right)
 \sin\!\left(\frac{\theta_{\max}y_f}{SL}g(x_f,t)\right),\\
 g(x_f,t)&=\sin\!\left(\omega t-\frac{2\pi W x_f}{BL}\right),
\end{aligned}
\label{eq:manta_kinematics}
\end{equation}

\noindent where $BL$ and $SL$ are the body-length and span, $W=BL/\lambda$ is the nondimensional chordwise wavenumber, and $\omega$ is the angular frequency. The deformation parameters are $\theta_{\max}=0.488$ and $k=0.960779$. The flow and acoustic calculations are based on $\mbox{Re}=3000$ and $\mathrm{Ma}=0.1$. The flow grid contains $425\times353\times353$ points. The acoustic domain is a cube with each side $35L$ long with $505\times425\times425$ grid points, and its target far-field spacing is chosen as $\lambda_{\mathrm{ac}}/10$.

\begin{figure*}[t]
 \centering
 \includegraphics[width=\textwidth]{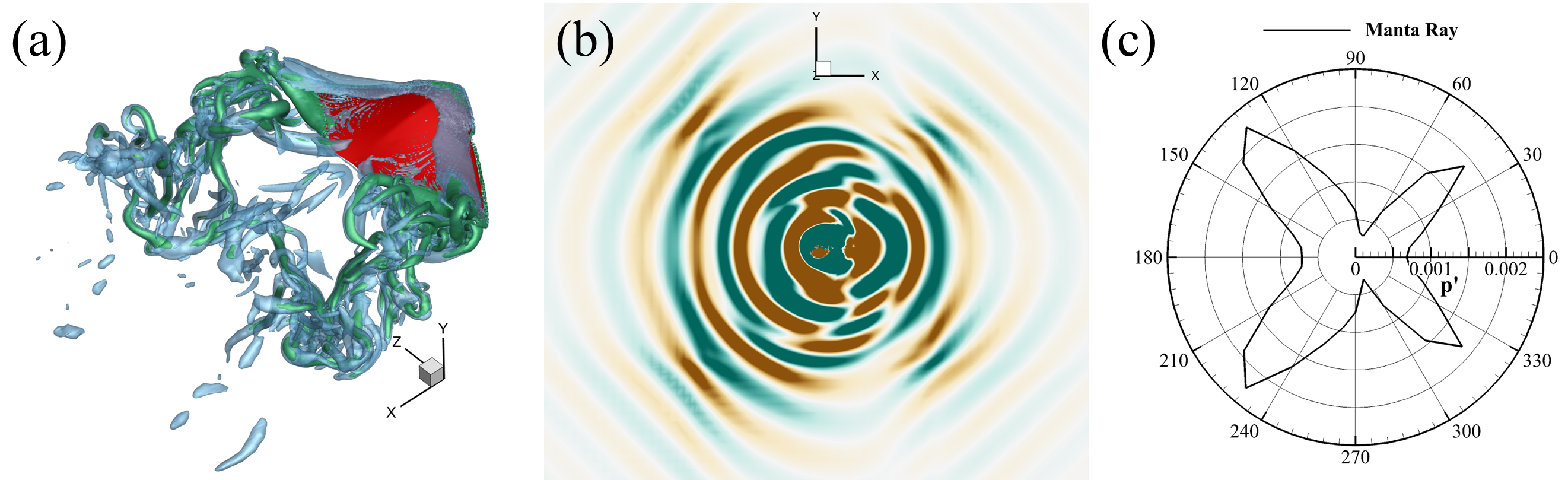}
 \caption{Manta-ray case: (a) instantaneous $Q$-criterion structures, with
 $Q=10$ in blue and $Q=60$ in green, (b) instantaneous nondimensional acoustic pressure on the $xy$- plane, and (c) far-field acoustic-pressure
 directivity.}
 \label{fig:Manta}
\end{figure*}

The computed flow contains connected vortical structures along the broad fins, near the body, and in the downstream wake (Fig.~\ref{fig:Manta}a). Despite this geometrical complexity, the acoustic pressure waves remain continuous as they propagate across the acoustic plane, while their azimuthally varying amplitude reflects the directional source distribution (Fig.~\ref{fig:Manta}b). The far-field signal forms four dominant oblique lobes, with weaker radiation in the streamwise and vertical directions  (Fig.~\ref{fig:Manta}c). This case demonstrates stable source transfer and wave propagation for a wide, highly deforming fin geometry.

\subsection{Harbor seal}
\label{sec:case_harbor}

The final application presented in this work considers the anatomically based harbor-seal model with dual hind flippers \cite{fardi2025characterizing}. The prescribed thunniform-like amplitude envelope, based on the reported swimming kinematics, is

\begin{equation}
 A(x)=0.02-0.1150x+0.2250x^2,
 \qquad 0<x<1,
 \label{eq:seal_envelope}
\end{equation}

The lateral motion is defined by:

\begin{equation}
 y(x,t)=A(x)\sin\!\left[2\pi
 \left(\frac{x}{\lambda^*}-ft\right)\right].
 \label{eq:seal_motion}
\end{equation}

\begin{figure*}[t]
 \centering
 \includegraphics[width=\textwidth]{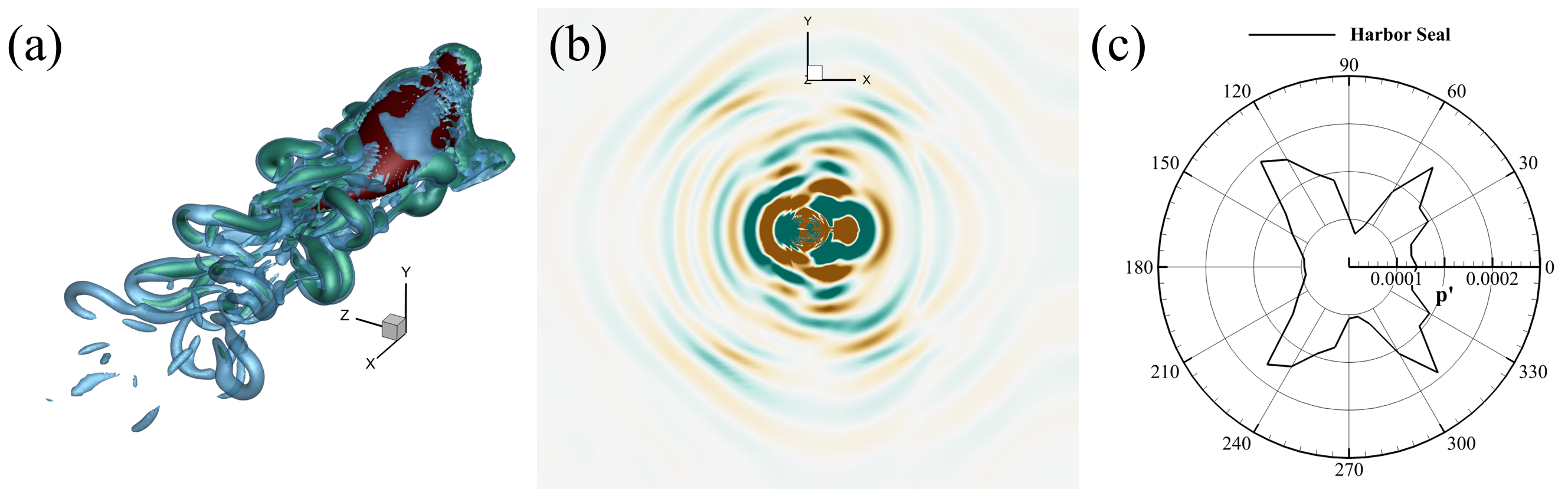}
 \caption{Harbor-seal case: (a) instantaneous $Q$-criterion structures, with
 $Q=10$ in blue and $Q=60$ in green; (b) instantaneous nondimensional acoustic
 pressure on the $x$--$y$ plane; and (c) far-field acoustic-pressure
 directivity.}
 \label{fig:Harbor}
\end{figure*}

The selected case uses $\lambda^*=1$, $\mbox{St}=0.35$, and $\mbox{Re}=3000$.
Figure~\ref{fig:Harbor}a shows that the method resolves vortices generated by the body and paired flippers together with the multi-scale downstream wake. The corresponding acoustic field consists of continuous outgoing waves with visible azimuthal modulation and near-field interference (Fig.~\ref{fig:Harbor}b). The directivity of the flow-induced sound is demonstrated by four principal oblique lobes and lower levels close to the streamwise directions (Fig.~\ref{fig:Harbor}c). Thus, the complete flow-acoustic procedure remains effective for a geometrically
complex body with two interacting propulsive appendages.

Taken together, the validation cases establish propagation, rigid-boundary, flow-induced-source, and moving-boundary accuracy, while the four applications extend the demonstration to realistic three-dimensional morphologies and a multiple-swimmer configuration. Across these cases, the dual-grid solver consistently resolves the near-body vortical source region, transfers the source to a larger acoustic domain, and produces stable outgoing pressure fields and physically distinct far-field directivities without requiring the flow grid to span the acoustic far field.

\section{Conclusions}
\label{sec:conclusions}
A fully parallel dual-grid framework has been developed for predicting flow-induced sound from complex moving and deforming bodies. The method combines an incompressible Navier-Stokes flow solver with an APE acoustic solver and applies a sharp-interface ghost-cell immersed boundary treatment on both grids. The flow and acoustic meshes are generated and decomposed independently, allowing each to be designed for its own physical scales. Near the body, matched and aligned grid points permit exact source transfer. Elsewhere, a precomputed bilinear or trilinear interpolation table and a fixed sparse MPI communication schedule transfer the flow pressure fluctuations to the acoustic grid. A Gaussian taper regularizes the source at the edge of the overlap region. Consequently, the finely resolved flow grid can remain restricted to the body and wake, whereas the acoustic grid can extend to the far field at a resolution selected from the acoustic wavelength. Four validation problems established the principal capabilities of the method. The convected Gaussian pulse reproduced the analytical pressure profile and left the domain without discernible reflection, verifying the high-order propagation scheme and far-field treatment. Scattering from a rigid circular cylinder agreed closely with the analytical pressure histories at both observation points, demonstrating accurate enforcement of the acoustic wall conditions. For flow past a cylinder at $Re=200$ and $\mathrm{Ma}=0.2$, the solver recovered the expected vortex street, lift-dipole radiation, wavelength, and transverse pressure decay reported in earlier studies. The traveling-wavy-foil case at $\mbox{Re}=1000$, $\mbox{St}=0.6$, and $\mathrm{Ma}=0.01$ reproduced the reference dipole pattern, the acoustic wavelength of approximately $33L$, and the phase and amplitude of the far-field waveform. Together, these tests verify the complete sequence from hydrodynamic-source prediction and cross-grid transfer to acoustic propagation around stationary and deforming immersed boundaries. The three-dimensional biological applications demonstrate the broader applicability of the proposed formulation. At otherwise comparable flow conditions, the eel and jack fish generate distinct wakes and radiation patterns, with the jack fish producing a substantially larger far-field pressure amplitude. The four-eel diamond formation produce constructive and destructive interference, increasing the pressure amplitude over most observation angles and creating a more strongly multi-lobed directivity than that of a single eel. The cases of a manta ray and a harbor seal retained continuous outgoing wave fields despite their broad deforming fins or interacting paired flippers, and both exhibited four dominant oblique radiation lobes with weaker streamwise radiation. These results show that morphology, gait, and arrangement of swimmers remain distinguishable from the near-body vortical field through to the acoustic far field. Hence, the presented framework here provides a flexible computational basis for studying the acoustic signatures of bio-inspired propulsors and other moving-boundary systems through one-way flow-acoustic coupling even for very low Mach numbers.


\section*{Acknowledgments}
MSU Khalid acknowledges the funding support from Natural Sciences and Engineering Research Council of Canada (NSERC) through the Discovery grant program. A. Fardi is thankful to Lakehead University for the graduate scholarship. The simulations reported in this work were performed on the supercomputing clusters administered and managed by the Digital Research Alliance of Canada. 


\bibliography{References}

@article{zhou2024effect,
  title={Effect of schooling on flow generated sounds from carangiform swimmers},
  author={Zhou, Ji and Seo, Jung-Hee and Mittal, Rajat},
  journal={Bioinspiration \& Biomimetics},
  volume={19},
  number={3},
  pages={036015},
  year={2024},
  publisher={IOP Publishing}
}

@article{williams1969sound,
  title={Sound generation by turbulence and surfaces in arbitrary motion},
  author={Williams, JE Ffowcs and Hawkings, David L},
  journal={Philosophical Transactions for the Royal Society of London. Series A, Mathematical and Physical Sciences},
  pages={321--342},
  year={1969},
  publisher={JSTOR}
}

@article{farassat1998acoustic,
  title={The acoustic analogy and the prediction of the noise of rotating blades},
  author={Farassat, Feri and Brentner, Kenneth S},
  journal={Theoretical and computational fluid dynamics},
  volume={10},
  number={1},
  pages={155--170},
  year={1998},
  publisher={Springer}
}

@article{lyrintzis2003surface,
  title={Surface integral methods in computational aeroacoustics—From the (CFD) near-field to the (Acoustic) far-field},
  author={Lyrintzis, Anastasios S},
  journal={International journal of aeroacoustics},
  volume={2},
  number={2},
  pages={95--128},
  year={2003},
  publisher={SAGE Publications Sage UK: London, England}
}

@article{colonius2004computational,
  title={Computational aeroacoustics: progress on nonlinear problems of sound generation},
  author={Colonius, Tim and Lele, Sanjiva K},
  journal={Progress in Aerospace sciences},
  volume={40},
  number={6},
  pages={345--416},
  year={2004},
  publisher={Elsevier}
}

@article{wang2006computational,
  title={Computational prediction of flow-generated sound},
  author={Wang, Meng and Freund, Jonathan B and Lele, Sanjiva K},
  journal={Annu. Rev. Fluid Mech.},
  volume={38},
  number={1},
  pages={483--512},
  year={2006},
  publisher={Annual Reviews}
}

@article{fish2020advantages,
  title={Advantages of aquatic animals as models for bio-inspired drones over present AUV technology},
  author={Fish, Frank E},
  journal={Bioinspiration \& biomimetics},
  volume={15},
  number={2},
  pages={025001},
  year={2020},
  publisher={IOP Publishing}
}

@article{fish2020bio,
  title={Bio-inspired aquatic drones: Overview},
  author={Fish, Frank E},
  journal={Bioinspiration \& Biomimetics},
  number={6},
  pages={060401},
  year={2020},
  publisher={IOP Publishing}
}

@article{marcin2020fish,
  title={Fish-like shaped robot for underwater surveillance and reconnaissance--Hull design and study of drag and noise},
  author={Marcin, Morawski and Adam, S{\l}ota and Jerzy, Zaj{\k{a}}c and Marcin, Malec},
  journal={Ocean Engineering},
  volume={217},
  pages={107889},
  year={2020},
  publisher={Elsevier}
}

@article{seo2021mosquitoes,
  title={Mosquitoes buzz and fruit flies don’ta comparative aeroacoustic analysis of wing-tone generation},
  author={Seo, Jung-Hee and Hedrick, Tyson L and Mittal, Rajat},
  journal={Bioinspiration \& Biomimetics},
  volume={16},
  number={4},
  pages={046019},
  year={2021},
  publisher={IOP Publishing}
}

@article{short2020influence,
  title={Influence of acoustics on the collective behaviour of a shoaling freshwater fish},
  author={Short, Matt and White, Paul R and Leighton, Timothy G and Kemp, Paul S},
  journal={Freshwater Biology},
  volume={65},
  number={12},
  pages={2186--2195},
  year={2020},
  publisher={Wiley Online Library}
}

@article{lighthill1952sound,
  title={On sound generated aerodynamically I. General theory},
  author={Lighthill, Michael James},
  journal={Proceedings of the Royal Society of London. Series A. Mathematical and Physical Sciences},
  volume={211},
  number={1107},
  pages={564--587},
  year={1952},
  publisher={The Royal Society London}
}

@article{curle1955influence,
  title={The influence of solid boundaries upon aerodynamic sound},
  author={Curle, Newby},
  journal={Proceedings of the Royal Society of London. Series A. Mathematical and Physical Sciences},
  volume={231},
  number={1187},
  pages={505--514},
  year={1955},
  publisher={The Royal Society London}
}

@article{farassat1988extension,
  title={Extension of Kirchhoff's formula to radiation from moving surfaces},
  author={Farassat, F and Myers, MK},
  journal={Journal of sound and vibration},
  volume={123},
  number={3},
  pages={451--460},
  year={1988},
  publisher={Elsevier}
}

@article{schoder2019hybrid,
  title={Hybrid aeroacoustic computations: State of art and new achievements},
  author={Schoder, Stefan and Kaltenbacher, Manfred},
  journal={Journal of Theoretical and Computational Acoustics},
  volume={27},
  number={04},
  pages={1950020},
  year={2019},
  publisher={World Scientific}
}

@article{hardin1994acoustic,
  title={An acoustic/viscous splitting technique for computational aeroacoustics},
  author={Hardin, Jay C and Pope, Dennis S},
  journal={Theoretical and computational fluid dynamics},
  volume={6},
  number={5},
  pages={323--340},
  year={1994},
  publisher={Springer}
}

@article{shen2004collocated,
  title={A collocated grid finite volume method for aeroacoustic computations of low-speed flows},
  author={Shen, Wen Zhong and Michelsen, Jess A and S{\o}rensen, Jens N{\o}rk{\ae}r},
  journal={Journal of Computational Physics},
  volume={196},
  number={1},
  pages={348--366},
  year={2004},
  publisher={Elsevier}
}

@article{shen1999aeroacoustic,
  title={Aeroacoustic modelling of low-speed flows},
  author={Shen, Wen Zhong and S{\o}rensen, Jens N{\o}rk{\ae}r},
  journal={Theoretical and Computational Fluid Dynamics},
  volume={13},
  number={4},
  pages={271--289},
  year={1999},
  publisher={Springer}
}

@article{ewert2003acoustic,
  title={Acoustic perturbation equations based on flow decomposition via source filtering},
  author={Ewert, Roland and Schr{\"o}der, Wolfgang},
  journal={Journal of Computational Physics},
  volume={188},
  number={2},
  pages={365--398},
  year={2003},
  publisher={Elsevier}
}

@article{seo2006linearized,
  title={Linearized perturbed compressible equations for low Mach number aeroacoustics},
  author={Seo, Jung H and Moon, Young J},
  journal={Journal of Computational Physics},
  volume={218},
  number={2},
  pages={702--719},
  year={2006},
  publisher={Elsevier}
}

@article{peskin1972flow,
  title={Flow patterns around heart valves: a numerical method},
  author={Peskin, Charles S},
  journal={Journal of computational physics},
  volume={10},
  number={2},
  pages={252--271},
  year={1972},
  publisher={Elsevier}
}

@article{mittal2005immersed,
  title={Immersed boundary methods},
  author={Mittal, Rajat and Iaccarino, Gianluca},
  journal={Annu. Rev. Fluid Mech.},
  volume={37},
  number={1},
  pages={239--261},
  year={2005},
  publisher={Annual Reviews}
}

@article{komatsu2016direct,
  title={Direct numerical simulation of aeroacoustic sound by volume penalization method},
  author={Komatsu, Ryu and Iwakami, Wakana and Hattori, Yuji},
  journal={Computers \& Fluids},
  volume={130},
  pages={24--36},
  year={2016},
  publisher={Elsevier}
}

@article{hattori2017mechanism,
  title={Mechanism of aeroacoustic sound generation and reduction in a flow past oscillating and fixed cylinders},
  author={Hattori, Yuji and Komatsu, Ryu},
  journal={Journal of Fluid Mechanics},
  volume={832},
  pages={241--268},
  year={2017},
  publisher={Cambridge University Press}
}

@article{wang2020immersed,
  title={An immersed boundary method for fluid--structure--acoustics interactions involving large deformations and complex geometries},
  author={Wang, Li and Tian, Fang-Bao and Lai, Joseph CS},
  journal={Journal of Fluids and Structures},
  volume={95},
  pages={102993},
  year={2020},
  publisher={Elsevier}
}

@article{cheng2021semi,
  title={A semi-implicit immersed boundary method for simulating viscous flow-induced sound with moving boundaries},
  author={Cheng, Long and Du, Lin and Wang, Xiaoyu and Sun, Xiaofeng and Tucker, Paul G},
  journal={Computer Methods in Applied Mechanics and Engineering},
  volume={373},
  pages={113438},
  year={2021},
  publisher={Elsevier}
}

@article{seo2011high,
  title={A high-order immersed boundary method for acoustic wave scattering and low-Mach number flow-induced sound in complex geometries},
  author={Seo, Jung Hee and Mittal, Rajat},
  journal={Journal of computational physics},
  volume={230},
  number={4},
  pages={1000--1019},
  year={2011},
  publisher={Elsevier}
}

@article{xie2020sharp,
  title={A sharp-interface Cartesian grid method for time-domain acoustic scattering from complex geometries},
  author={Xie, Fangtao and Qu, Yegao and Islam, Md Ariful and Meng, Guang},
  journal={Computers \& Fluids},
  volume={202},
  pages={104498},
  year={2020},
  publisher={Elsevier}
}

@article{he2022improved,
  title={An improved hydrodynamic/acoustic splitting method for fluid--structure interaction feedback with elastic boundaries},
  author={He, Yanfei and Zhang, Xingwu and Zhang, Tao and Geng, Jia and Liu, Jinxin and Chen, Xuefeng},
  journal={Physics of Fluids},
  volume={34},
  number={2},
  year={2022},
  publisher={AIP Publishing}
}

@article{zhao2021sharp,
  title={A sharp interface immersed boundary method for flow-induced noise prediction using acoustic perturbation equations},
  author={Zhao, Cheng and Yang, Yan and Zhang, Tao and Dong, Haibo and Hou, Guoxiang},
  journal={Computers \& Fluids},
  volume={227},
  pages={105032},
  year={2021},
  publisher={Elsevier}
}

@article{zhao2024hybrid,
  title={Hybrid approach for simulating flow-induced sound around moving bodies based on ghost-cell immersed boundary method},
  author={Zhao, Cheng and Li, Hong-Gang and Li, Xue-Gang and Yang, Yan and Cui, Kai},
  journal={Acta Mechanica Sinica},
  volume={40},
  number={12},
  pages={323621},
  year={2024},
  publisher={Springer}
}

@article{smith2022hybrid,
  title={A hybrid computational aeroacoustic model with application to turbulent flows over foil and bluff bodies},
  author={Smith, Tom A and Ventikos, Yiannis},
  journal={Journal of Sound and Vibration},
  volume={526},
  pages={116773},
  year={2022},
  publisher={Elsevier}
}

@article{purohit2014numerical,
  title={A numerical investigation on effects of structural flexibility on aerodynamic far field sound},
  author={Purohit, Ashish and Darpe, Ashish K and Singh, SP},
  journal={Computers \& Fluids},
  volume={89},
  pages={143--152},
  year={2014},
  publisher={Elsevier}
}

@article{labbe2013cfd,
  title={A CFD/CAA coupling method applied to jet noise prediction},
  author={Labbe, OPCRGHM and Peyret, C and Rahier, G and Huet, M},
  journal={Computers \& Fluids},
  volume={86},
  pages={1--13},
  year={2013},
  publisher={Elsevier}
}

@article{groschel2008noise,
  title={Noise prediction for a turbulent jet using different hybrid methods},
  author={Gr{\"o}schel, E and Schr{\"o}der, W and Renze, P and Meinke, M and Comte, P},
  journal={Computers \& fluids},
  volume={37},
  number={4},
  pages={414--426},
  year={2008},
  publisher={Elsevier}
}

@article{moon2010hybrid,
  title={A hybrid prediction method for low-subsonic turbulent flow noise},
  author={Moon, YJ and Seo, JH and Bae, YM and Roger, Michel and Becker, St{\'e}phanie},
  journal={Computers \& Fluids},
  volume={39},
  number={7},
  pages={1125--1135},
  year={2010},
  publisher={Elsevier}
}

@article{farooq4874977accurate,
  title={An Accurate Immersed Boundary Method Using Radial-Basis Functions for Incompressible Flows},
  author={Farooq, Hamayun and Akhtar, Imran and Hemmati, Arman and Khalid, Muhammad Saif Ullah},
  journal={Under Review, Available at SSRN 4874977},
  year={2024},
  publisher={}
}

@article{mittal2008versatile,
  title={A versatile sharp interface immersed boundary method for incompressible flows with complex boundaries},
  author={Mittal, Rajat and Dong, Haibo and Bozkurttas, Meliha and Najjar, FM and Vargas, Abel and Von Loebbecke, Alfred},
  journal={Journal of computational physics},
  volume={227},
  number={10},
  pages={4825--4852},
  year={2008},
  publisher={Elsevier}
}

@article{tam1993dispersion,
  title={Dispersion-relation-preserving finite difference schemes for computational acoustics},
  author={Tam, Christopher KW and Webb, Jay C},
  journal={Journal of computational physics},
  volume={107},
  number={2},
  pages={262--281},
  year={1993},
  publisher={Elsevier}
}

@article{tam1995computational,
  title={Computational aeroacoustics-Issues and methods},
  author={Tam, Christopher KW},
  journal={AIAA journal},
  volume={33},
  number={10},
  pages={1788--1796},
  year={1995}
}

@article{hu1996low,
  title={Low-dissipation and low-dispersion Runge--Kutta schemes for computational acoustics},
  author={Hu, FQ and Hussaini, M Yousuff and Manthey, JL},
  journal={Journal of computational physics},
  volume={124},
  number={1},
  pages={177--191},
  year={1996},
  publisher={Elsevier}
}

@article{bogey2004family,
  title={A family of low dispersive and low dissipative explicit schemes for flow and noise computations},
  author={Bogey, Christophe and Bailly, Christophe},
  journal={Journal of Computational physics},
  volume={194},
  number={1},
  pages={194--214},
  year={2004},
  publisher={Elsevier}
}

@inproceedings{edgar2003general,
  title={A general buffer zone-type non-reflecting boundary condition for computational aeroacoustics},
  author={Edgar, Nathan and Visbal, Miguel},
  booktitle={9th AIAA/CEAS Aeroacoustics Conference and Exhibit},
  pages={3300},
  year={2003}
}

@inproceedings{hardin1994icase,
  title={ICASE/LaRC workshop on benchmark problems in computational aeroacoustics},
  author={Hardin, JC and Ristorcelli, JR and Tam, CKW},
  booktitle={NASA conference publication},
  volume={3300},
  year={1994}
}

@book{tam1997second,
  title={Second computational aeroacoustics (CAA) workshop on benchmark problems},
  author={Tam, Christopher KW and Hardin, Jay C},
  volume={3352},
  year={1997},
  publisher={National Aeronautics and Space Administration, Langley Research Center}
}

@article{zhao2022sound,
  title={Sound generated by flow over two traveling wavy foils in a side-by-side arrangement},
  author={Zhao, Cheng and Zhang, Tao and Yang, Yan and Dong, Haibo},
  journal={Physics of Fluids},
  volume={34},
  number={12},
  year={2022},
  publisher={AIP Publishing}
}

@article{zhang2022vortex,
  title={Vortex dynamics and hydrodynamic performance enhancement mechanism in batoid fish oscillatory swimming},
  author={Zhang, Dong and Huang, Qiao-Gao and Pan, Guang and Yang, Li-Ming and Huang, Wei-Xi},
  journal={Journal of Fluid Mechanics},
  volume={930},
  pages={A28},
  year={2022},
  publisher={Cambridge University Press}
}

@article{fardi2025characterizing,
  title={Characterizing the role of hind flippers in hydrodynamics of a harbor seal},
  author={Fardi, Amirhossein and Farooq, Hamayun and Akhtar, Imran and Hemmati, Arman and Saif Ullah Khalid, Muhammad},
  journal={Bioinspiration \& Biomimetics},
  volume={20},
  number={4},
  pages={046010},
  year={2025},
  publisher={IOP Publishing}
}

@article{khalid2021anguilliform,
	title={Why do anguilliform swimmers perform undulation with wavelengths shorter than their bodylengths?},
	author={Khalid, Muhammad Saif Ullah and Wang, Junshi and Akhtar, Imran and Dong, Haibo and Liu, Moubin and Hemmati, Arman},
	journal={Physics of Fluids},
	volume={33},
	number={3},
	year={2021},
	publisher={AIP Publishing}
}

@article{khalid2021larger,
  title={Larger wavelengths suit hydrodynamics of carangiform swimmers},
  author={Khalid, Muhammad Saif Ullah and Wang, Junshi and Akhtar, Imran and Dong, Haibo and Liu, Moubin and Hemmati, Arman},
  journal={Physical Review Fluids},
  volume={6},
  number={7},
  pages={073101},
  year={2021},
  publisher={APS}
}

@article{ligman2024comprehensive,
  title={A comprehensive review of hydrodynamic studies on fish schooling},
  author={Ligman, Montana and Lund, Joshua and F{\"u}rth, Mirjam},
  journal={Bioinspiration \& Biomimetics},
  volume={19},
  number={1},
  pages={011002},
  year={2024},
  publisher={IOP Publishing}
}

\end{document}